\documentclass{aa}  
\usepackage{booktabs}
\usepackage{txfonts}
\usepackage{graphicx,amsmath,amssymb,multirow}
\usepackage{xspace}
\usepackage{color}
\usepackage{natbib}
\usepackage{nicefrac}
\usepackage{ulem}
\usepackage{rotating}
\usepackage{longtable}
\usepackage{multicol}
\usepackage{supertabular}

\newcommand{\Jband}{{\it J}\xspace}

\newcommand{\eg}{e.g.\xspace}
\newcommand{\sn}{SN\xspace}
\newcommand{\sne}{SNe\xspace}
\newcommand{\snia}{SN~Ia\xspace}
\newcommand{\sneia}{SNe~Ia\xspace}

\def\msun{M_\odot}

\begin{document}

	\title{Searching for supernovae in the multiply-imaged galaxies behind the gravitational telescope A370\thanks{Based on observations made with  European Southern
			Observatory (ESO) telescopes at the Paranal Observatory under programme ID 091.A-0108 and ID 093.A-0278  PI: A. Goobar.}}

\titlerunning{Search for lensed SNe behind A370}
\authorrunning{Petrushevska et al.}
	\subtitle{}

	\author{T.~Petrushevska\inst{1,2} \and 
				A.~Goobar\inst{1} \and
				D. J. Lagattuta \inst{3} \and
		R.~Amanullah\inst{1} \and 
		L.~Hangard\inst{1} \and
		S.~Fabbro\inst{4} \and
		C. Lidman\inst{5} \and
		K. Paech\inst{6,7} \and
		J. Richard\inst{3} \and
		J.~P.~Kneib\inst{8}		
	}
	\institute{Oskar Klein Centre, Department of Physics, Stockholm University, SE 106 91 Stockholm, Sweden 
		\and{Centre for Astrophysics and Cosmology, University of Nova Gorica, Vipavska 11c, 5270 Ajdov\v{s}\u{c}ina, Slovenia}\\
		\email{tanja.petrushevska@ung.si} 
		\and Univ Lyon, Univ Lyon1, Ens de Lyon, CNRS, Centre de Recherche Astrophysique de Lyon UMR5574, F-69230, Saint-Genis-Laval, France 
		\and NRC Herzberg Institute for Astrophysics, 5071 West Saanich Road, Victoria V9E 2E7, British Columbia, Canada
		\and Australian Astronomical Observatory, 105 Delhi Road, North Ryde NSW, 2113, Australia
		\and Universit\"ats-Sternwarte, Fakult\"at f\"ur Physik, Ludwig-Maximilians Universitaet M\"unchen, Scheinerstr. 1, D-81679 Muenchen, Germany
		\and Excellence Cluster Universe, Boltzmannstr. 2, D-85748 Garching, Germany
		\and Laboratoire d'Astrophysique, Ecole Polytechnique F\'ed\'erale de Lausanne (EPFL), Observatoire de Sauverny, CH-1290 Versoix, Switzerland}
	
	\date{Received 12/07/2017; accepted 26/02/2018}

	\abstract
	{}
	{ Strong lensing by massive galaxy clusters can provide magnification of the flux and even multiple images of the galaxies that lie behind them. This phenomenon facilitates observations of high-redshift supernovae (SNe), that would otherwise remain undetected. Type Ia supernovae (SNe Ia) detections are of particular interest because of their standard brightness, since they can be used to improve either cluster lensing models or cosmological parameter measurements.}
	{We present a ground-based, near-infrared search for lensed SNe behind the galaxy cluster Abell 370. Our survey was based on 15  epochs of $J$-band observations with the HAWK-I instrument on the Very Large Telescope (VLT). We use Hubble Space Telescope (HST) photometry to infer the global properties of the multiply-imaged galaxies. Using a recently published lensing model of Abell 370, we also present the predicted magnifications and time delays between the images.}
	{ In our survey, we did not discover any live SNe from the 13 lensed galaxies with 47 multiple images behind Abell 370. This is consistent with the expectation of $0.09 \pm 0.02$ SNe calculated based on the measured star formation rate. We compare the expectations of discovering strongly lensed SNe in our survey and that performed with HST during the Hubble Frontier Fields (HFF) programme. We also show the expectations of search campaigns that can be conducted with future facilities, such as the James Webb Space Telescope (JWST) or the Wide-Field Infrared Survey Telescope (WFIRST). We show that the NIRCam instrument aboard the JWST will be sensitive to most SN multiple images in the strongly lensed galaxies and thus will be able to measure their time delays if observations are scheduled accordingly.}
	{}
	
	\maketitle

	\section{Introduction}
	\citet{1937ApJ....86..217Z} suggested that massive astrophysical objects can act as gravitational lenses, magnifying  and producing multiple images of the same background source.
	A good example of a gravitational lens is  Abell 370 (A370), one of the most massive and well-studied galaxy clusters. A370 lensing properties allowed the discovery of one of the first extended-source lensed galaxies (the Giant arc; \citealt{1987A&A...172L..14S}). In a more recent study, the A370 mass distribution was modelled with Hubble Frontier Fields (HFF) images and GLASS spectroscopy \citep{2017MNRAS.473.4279D}. These authors identified  dozens  of multiple images of $13$ strongly lensed background galaxy sources behind A370. \citet{2017MNRAS.469.3946L} (L17) identified nine new multiply-imaged systems and secured their redshifts with the help of Multi-Unit Spectroscopic Explorer (MUSE) spectroscopy and archival Hubble Space Telescope (HST) images. In total, there are 22 known unique systems with at least 69 individual images. Using these new constraints, L17 updated the lens model of A370, probing the mass distribution from cluster to galaxy scales. 
	
	High-$z$ galaxies that appear as multiple images can host supernova (SN) explosions. Owing to the magnification boost due to lensing, SNe that would normally be too faint to be detected by current telescopes, can be observed. As the light from the multiple images travels on different paths, the arrival of SN images exhibits a time delay. The time delay depends on the Hubble parameter, and to a lesser degree, other cosmological parameters  \citep{2001ApJ...556L..71H,2002A&A...393...25G,2003ApJ...592...17B, 2003MNRAS.338L..25O,2007ApJ...660....1O}.
	There is also a gravitational dependence on the time delay, i.e. from the lensing potential. Therefore, strongly lensed SNe can be used as tools to examine both global cosmology and the local environment of the cluster lenses \citep[see \eg][]{Third}. 
	
   The first multiply-imaged \sn at $z=1.489$ (dubbed SN Refsdal) was discovered behind the galaxy cluster MACS J1149.6+2223 \citep{2015Sci...347.1123K}.  The image of SN Refsdal re-appeared almost a year later in another host galaxy image, as predicted independently by several lensing models, thus providing a test of the accuracy of the lens model predictions \citep{2016ApJ...822...78G, 2016ApJ...819L...8K, 2016MNRAS.457.2029J,2016MNRAS.456..356D}. It was identified as a core-collapse (CC) type of explosion \citep{2016ApJ...831..205K}. 
	
	 Because of their standard candle nature, observations of  \sneia through lensing clusters are even more interesting. By estimating the absolute magnification of \sneia, it is possible to break the so-called mass-sheet degeneracy of gravitational lenses \citep{2003MNRAS.338L..25O}. Thus, even single magnified \snia could be used to put constraints on the lensing potential, if the background cosmology is assumed to be known \citep{2014ApJ...786....9P,2014MNRAS.440.2742N, 2015ApJ...811...70R}. This can be further improved by observations of strongly lensed \sneia, through the measurements of time delays between the multiple images. In contrast, if the lensing potential is well known, measurements of time delays of any transient source can be used to measure cosmological parameters. 
	
 The first resolved multiply-imaged SN~Ia, iPTF16geu \citep{2017Sci...356..291G}, was magnified by a galaxy lens, rather than a galaxy cluster. For this reason, the expected time delays were about one day, a timescale that is difficult to probe with SNe~Ia light curves. Cluster lensing timescales are typically much longer, which makes their measurement potentially more feasible  \citep{2017arXiv171205800V}, especially if the lens potential is well studied and the predicted time delays have small uncertainties.
	
A dedicated ground-based search for SNe behind Abell~1689 was presented in 	\citet[P16 hereafter]{2016A&A...594A..54P}, after the pilot studies carried out by \citet{2009A&A...507...61S}, \citet{Second} and \citet{2011ApJ...742L...7A}. We present a similar near-infrared (NIR) survey of A370. 

Throughout this paper the cosmology $\Omega_{\Lambda}=0.7$, $\Omega_{\rm M}=0.3$ and $H_0=70$ is assumed and all presented magnitudes are in the Vega system.

	\section{A370 HAWK-I observations and transient search}
	All data presented were obtained with the High Acuity Wide field K-band imager  (HAWK-I; \citealt{2004SPIE.5492.1763P,2006SPIE.6269E..0WC}) mounted on the VLT (Programmes ID 090.A-0492, 091.A-0108, P.I. Goobar). HAWK-I has a combined  $7.5' \times 7.5'$ field of view, divided into four equally-sized pixel arrays arranged in a square grid.  Since the core of A370 covers  a smaller area than a single HAWK-I detector, we placed the cluster at the centre of one of the chips for each exposure.  To optimise the survey for finding high-$z$ SNe, all observations are carried out using the \Jband-band filter, covering the wavelength range $1.17 - 1.34$  $\mu m$.  Nearby SN spectra peak in the optical region of the spectrum, but at high-$z$, their light is redshifted to the NIR. 
	Furthermore, considering the duration of light curves of high-$z$ and the effects of time dilation, we kept $\sim1$ month cadence while the galaxy cluster was visible on the sky. The data were taken over 16 separate nights in 2012 and 2013, resulting in a total exposure time of 18 hrs and 22 min, and a median limiting magnitude of $m_J = 23.6$.  Details of the observations are presented in Table~\ref{HAWKI_data}. The observations were executed in blocks of 29 or 58 exposures made of six 20 second integrations.  To allow us to perform accurate sky subtraction, the telescope was offset in a semi-random manner between each exposure. We reduced the data following the procedure presented in P16. A full description can be found
	in that work and we only briefly describe the data reduction process. The individual frames of each epoch were dark subtracted, flat fielded and the pixels that were saturated or affected by cosmic rays were masked. A sky-frame was constructed for each exposure using standard NIR imaging reduction techniques;  the only exception was that we took advantage of the fact that the same field was observed repeatedly and therefore built an object mask that was updated as more data became available.
	 A sky-subtraction and geometrical distortion correction was performed. Then the frames were geometrically aligned and combined.

 After the new image was subtracted from the reference image, a \sn candidate detection algorithm was run on the difference image. In Appendix~\ref{sec:patches}, image (new and reference) and subtraction stamps are shown with zoomed-in on the multiply-imaged galaxies.
 We performed a transient search on each image immediately after it was obtained and reduced (see section 3 of P16 for a full description of this process). Additionally, after the survey was completed we repeated the search using the full, combined data set. In this post-survey search, epochs that were acquired less than 14 days apart were co-added to obtain better image depth. After combining the images, we were left with seven stacks in total (see Table~\ref{HAWKI_data}). 
 For the purposes of the transient search, we also separated each stack into two \textit{substacks}, where only half of the frames are combined. 
 
 We limited our search criteria to include only objects with a S/N~$\ge4$ in the subtracted image.  This limit applied to both the stacks and substacks, where the object needed to be present in both substack images with S/N~$\ge4$. There are total of 1753 resulting candidates that were inspected meticulously by eye and ranked by members of our team independently according to the procedure presented in P16. Upon completing this inspection, none of the candidates were classified as probable SNe.
Most of the candidates appear in only a single search epoch and were discarded for various reasons, for example subtraction artifacts at the cores of bright galaxies, AGN sources and cosmic rays.  By requiring candidates to appear in at least two consecutive epochs instead, the number of candidates reduces to 314 candidates. 

	\begin{table}
		\begin{center}
			\caption{A370 observations with VLT/HAWK-I in the \Jband band. \label{HAWKI_data}}
			\begin{tabular}{lcccc}
\toprule
				Date & Exp. time & Seeing   &	$m_{lim}^a$  \\

				& (sec) & (arcsec) &  (mag)  \\
				\midrule
				2012-10-13&3480& 0.67&\\
				2012-10-14&3480&0.45&used as ref. \\
								\midrule
				2012-11-06&3480&0.44 &\\
				2012-11-09&6960& 0.40&\\
			    2012-11-10&6960&0.39& \\
				2012-11-11&3480& 0.56&23.46\\
							\midrule
				2012-12-09&3480&  0.57&\\
				2012-12-13&3480&0.50 &23.40\\
							\midrule
				2012-12-31&3480&0.40& \\
				2013-01-06&3480&0.69 &23.83\\
							\midrule

				2013-08-02&3480&0.76 &23.16\\
						\midrule
					2013-08-28 &3480&0.37&24.04\\
						\midrule

				2013-10-24&3480&0.57& \\
				2013-10-28&3480&0.56&23.77\\
						\midrule
				2013-11-18&3480& 0.67&\\
				2013-11-28&6960&0.52&23.59 \\
			  \bottomrule
			  \end{tabular}
				\tablefoot{					The line indicates that those observations were combined for a deeper image.  \\
					$^a$ $m_{lim}$ indicates $5\sigma$ limiting depth for the combined image.}
		\end{center}
	\end{table}

\section{Supernovae in the multiply-imaged galaxies behind A370}
\label{sec:multi}
\begin{figure*} [htbp]
	\begin{center}
	\includegraphics[width=0.8\textwidth]{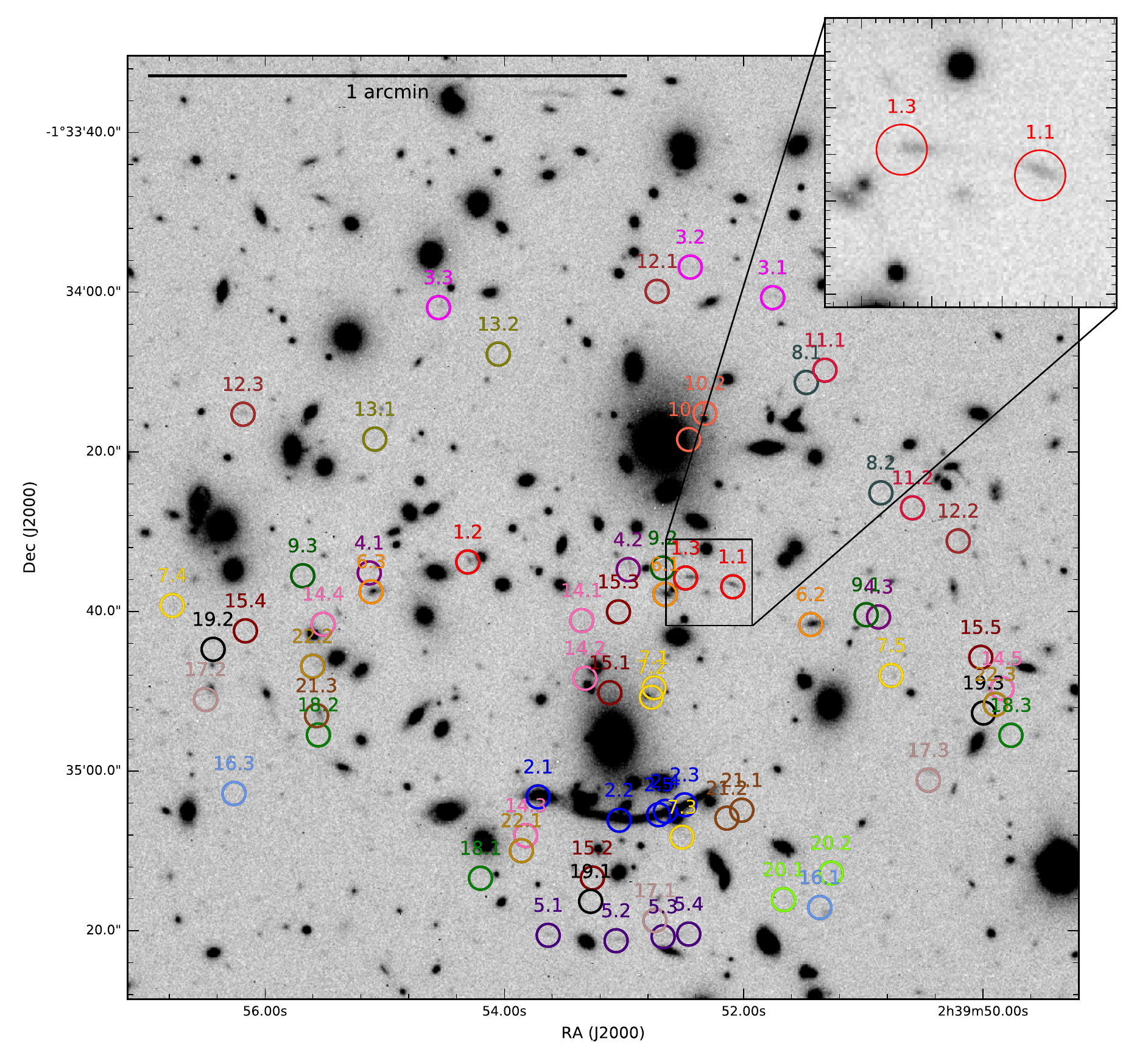}
	\caption{Cropped J-band VLT/HAWK-I image overplotted with the positions of all the 22 background galaxies with 69 multiple images that were presented in \citet{2017MNRAS.469.3946L}. The images that belong to the same system are shown with the same colour. There are 13 systems that have a secured redshift, which means that is at least one image in the system with spectroscopic redshift. The predicted magnifications for these galaxies from the lensing model are approximately $\sim 1-4$ magnitudes (see Table~\ref{table:multi}). As an example, a  zoomed-in view of multiply-imaged galaxies 1.3 and 1.1 are shown, which are magnified by $2.81\pm0.08$ and $2.80\pm0.07$ mag, respectively. }
		\label{multiple}
	\end{center}
\end{figure*}

In this section, we calculate the expected number of SNe in the multiply-lensed galaxies in our HAWK-I survey. We only consider the multiple image systems that have secured redshifts, where at least one of the images in the system has a spectroscopic redshift. We do not include those with only photometric redshifts, which would introduce an additional source of error. There are 13 out of the total 22 systems that satisfy this criteria. These 13 systems are composed of 47 individual multiple images whose properties are presented in Table~\ref{table:multi} and shown in Figure~\ref{multiple}. 

\begin{figure*}[htbp]
	\begin{center}
		\includegraphics[width=\textwidth]{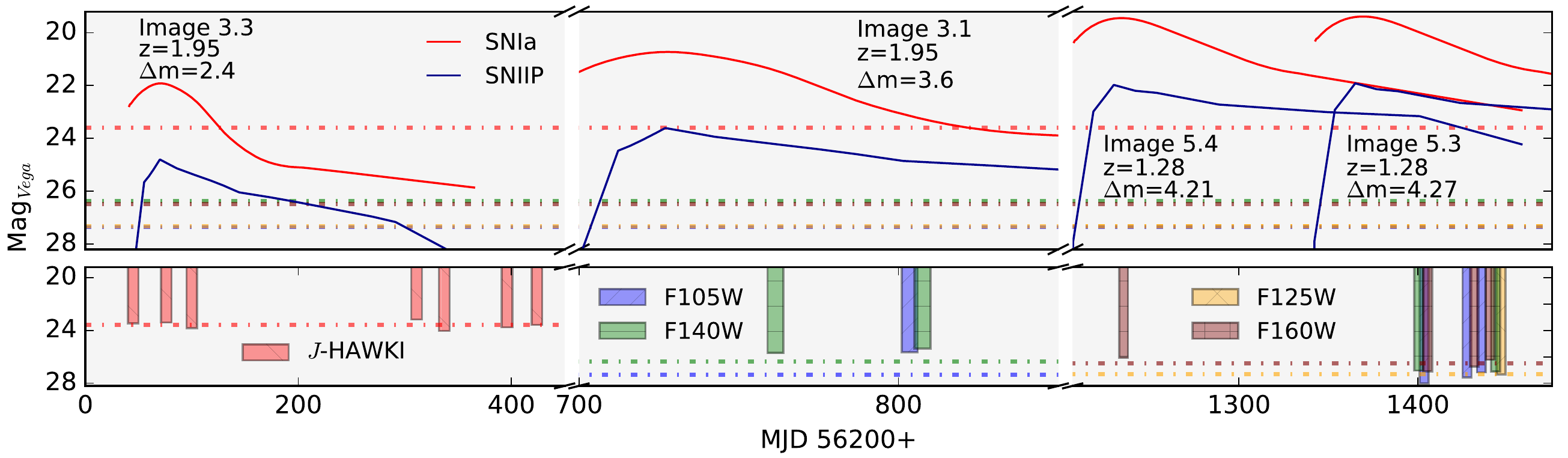}
		\caption{Top panels: Simulated light curves of SNe Ia and IIP in F125W band for pairs of images of strongly lensed galaxies behind A370. The redshift and magnifications of the galaxy images are shown. If a SN had exploded in galaxy image 3.3, another image would have been observed in galaxy 3.1 with a time delay of $\sim660$ days. Similarly, if it had exploded in 5.4, the SN would have reappeared in galaxy image 5.3 with a time delay of $\sim130$ days.
			Bottom panels: Magnitude limit for the HAWK-I in \Jband band and HFF observations  in F105W, F125W, F140W and F160W bands of A370 as presented in Tables \ref{HAWKI_data} and \ref{HFF_data1}.
			Both panels: The dotted lines indicate the median magnitude limit for the HAWK-I and HFF/WFC3 surveys. }
		\label{maglimit}
	\end{center}
\end{figure*}

Following P16, we calculated the expected number of \sne in each galaxy $N_i$, by multiplying the \sn rate $R_i$ and the control time $T_i$,
\begin{equation}
N_i=R_i\cdot T_i,
\end{equation}
where $i$ indicates the individual galaxies. The control time above the detection threshold for a \sn of type $j$, $T_j$, is a function of the \sn light curve, absolute intrinsic SN brightness, detection efficiency, extinction by dust, and lensing magnification. The probability distributions of the absolute intrinsic brightness for the SN types are assumed to be Gaussian with values taken from \citet{2014AJ....147..118R}. 
We calculated the control time for \sneia and CC separately. The control time depends on the properties of the light curves, therefore different subtypes of CC~\sne have different control times. The total CC control time was obtained by weighting the contribution from the various CC~\sn subtypes with their fractions from \citet{Li} and then summed.  We obtained the final number of SNe by summing over the individual estimates from each multiple image in the system.  Doing this, we were able to determine the sensitivity of our survey to SNe exploding in the galaxies with multiple images. For SNe~Ia, the survey was sensitive to 29 galaxy images originating from 9 unique systems (1, 2, 3, 4, 5, 6, 7, 9 and 21), with redshifts ranging between $0.72<z<2.75$. Considering CC~SNe, for example type IIP, which are the most common type of CC~SNe in the local universe, the survey was sensitive to 16 galaxy images coming from 5 unique systems (1, 2, 3, 5,  and 21), with redshifts ranging between $0.72<z<1.95$. SNe IIP are on average intrinsically fainter than SNe~Ia, thus they are only visible in the galaxies at lower redshift and with high magnification. A few examples of simulated light curves that could have been observed with the HAWK-I survey are shown in the top panel of Figure~\ref{maglimit}.  The synthetic light curves in the observer filters for redshift $z$ were obtained by applying cross-filter k-corrections \citep{1996PASP..108..190K}. The mean absolute magnitudes were adopted from \citet{2014AJ....147..118R}. Furthermore,  the lensing magnification from the galaxy cluster is also taken into consideration.

To obtain the expected SN rate in each galaxy, $R_i$,  the relationships between star formation rate (SFR) and SN rate was used. We estimated the ongoing SFR for each galaxy with the help of Fitting and Analysis of Spectral Templates (FAST)\footnote{http://w.astro.berkeley.edu/~mariska/FAST.html}. The FAST code fits stellar populations models to broadband photometry, providing the global properties of galaxies such as stellar mass, age and SFR (described in the appendix of \citealt{2009ApJ...700..221K}). We chose FAST as it has been extensively and successfully used to study large sets of galaxies, even at high redshifts \citep[see~e.g.][]{2009ApJ...705L..71K,2017ApJ...835..281M,2017ApJ...835..254M}. We measured the photometry of our galaxy sample using archival observations from HST. Most of these were taken as part of the HFF\footnote{https://archive.stsci.edu/prepds/frontier/abell370.html} project with the Advanced Camera for Surveys (ACS) in three optical bands, F435W, F606W, and F814W and with the Wide-Field Camera 3 (WFC3) in four NIR bands: F105W, F125W, F140W, and F160W.  These imaging data represent one of the deepest, highest resolution data with average limiting magnitude  27.4 in the optical bands and 26.8 in the NIR bands.  Photometric estimates in each band are obtained using Source Extractor (SExtractor; \citealt{1996A&AS..117..393B}). The light from the brightest cluster galaxies was subtracted away following the procedure in L17, to obtain less contaminated photometry of the background sources. We note that this procedure tends to increase the error on the final magnitude. The HST photometry for all systems is shown in Table~\ref{photometry}.  

To obtain the intrinsic brightness of the galaxies, the predicted magnification from the galaxy cluster must be taken in to account. This means that our estimate of the brightness of the galaxies depends on the assumed lensing model. In Table~\ref{table:multi}, we show the magnifications from the galaxy cluster which we compute from the A370 lens model in L17. 
To obtain an estimate of total magnification, we use the entire surface of a given galaxy, instead of simply adopting the value measured at the object's centroid.  Specifically, we calculate an individual magnification for each pixel associated with the object in the HST data and then take the median value of the entire map.  To identify these pixels, we use the SExtractor segmentation map generated from our photometry measurements. This process produces a more robust magnification estimate compared to	the centroid approach, as it accounts for magnification gradients that naturally occur in extended sources.  Additionally, we prefer a median	estimator over a simple mean, as it is less sensitive to extreme magnification values that occur near the lensing critical curve.  This	is especially important in highly distorted arcs and merging pairs, such as Systems 2 and 3. After applying these magnification corrections to the galaxies'	observed magnitudes (Table~\ref{photometry}), we find that the corrected intrinsic brightness of nearly all images of a given lensed galaxy agree with each other within the flux errors, as we would expect.  The exceptions to this are merging pairs of images which by definition only contain a fraction of the total galaxy.  This affects images in Systems 2 (2.2 through 2.5), 3 (3.1 and 3.2), 5 (5.3 and 5.4), 7 (7.1 and 7.2), and 21 (21.1 and 21.2). 

As the progenitors of CC~SNe are short-lived stars, it is expected that they trace ongoing SFR. Therefore, the CC~SN rate for each galaxy in units of $yr^{-1}$ can be estimated from the relation:
\begin{equation}
R_{\rm CC} = k_8^{50} \cdot \rm SFR,
\end{equation} 
where the SFR is given in in units $\rm \msun$yr$^{-1}$, and $k_8^{50}$ is the scale factor, given as the number of stars per unit mass that explode as CC SNe. The scale factor depends on the initial mass function (IMF) and the range of stellar masses that explode as CC SNe. As in P16,  we used a \cite{1955ApJ...121..161S} IMF and $\rm M_{min} = 8 \, \rm \msun$ and $M_{max} = 50\, \msun$, which results in scale factor of $k_8^{50}=0.007 \,\rm\msun^{-1}$. The \snia rates were obtained from the \citet{2005ApJ...629L..85S} two-component model
\begin{equation}
R_{Ia} = A \cdot {\rm SFR} + B \cdot M_*,
\end{equation} 
where $A$ and $B$ are constants, determined from observations for which we use literature values from \citet{2012ApJ...755...61S}, while the stellar mass $\rm M_*$ of the individual galaxies was taken from the FAST best fits. The total expected number of SNe over all the systems are then simply summed. The result is  $N_{\rm CC}=0.07\pm0.02$ and   $N_{\rm Ia}=0.02 \pm0.01 $ 
for CC~SNe and  \sneia, respectively, which is consistent with our observations.

\begin{table*}

		\centering
	\caption{Properties of the multiply-imaged galaxies behind A370. Time delays ($\Delta t$) of the multiply-imaged galaxies as predicted from the lensing model in L17 are listed in column 3. The time delays are measured relative to the first image in column 1.  The reference image in each case is that with the shortest path length between the galaxy and us. The time delays are with respect to the reference image. The magnification is given in magnitudes in column 4 and 5, and is computed using the lensing model in L17. Estimates of the core-collapse and SN Ia rates are given with $1\sigma$ errors in column 6 and 7, respectively. Given that the SFR of the galaxy depends on the luminosity of the galaxy, it is dependent of the predicted magnification, thus the lensing model. In columns 8-14, the detectability of a lensed SN for the different SN types assuming a 27.5 mag in F150W, e.g. in a survey using 1 hour exposure of the JWST/NIRCam instrument. In order to be observable, the luminosity of both images must lie above the observation threshold.}
	\label{table:multi}
	\small{
	\begin{tabular}{l c c r r r r c c c c c c c}
\toprule
			Images$^{\rm a}$ & $z$ & \multicolumn{1}{c}{$\Delta t$} & \multicolumn{1}{c}{$\Delta m_1$} & \multicolumn{1}{c}{$\Delta m_2$} & \multicolumn{1}{c}{$r_{\rm CC}$ $\times$  $10^{2}$} & \multicolumn{1}{c}{$r_{\rm Ia}$ $\times$ $10^{3}$} &  \multicolumn{7}{c}{Detectable SN types} \\
		& & \multicolumn{1}{c}{(d)} & \multicolumn{2}{c}{(mag)} &  \multicolumn{2}{c}{(yr$^{-1}$)} & Ia & IIP & IIL & IIn & Ib & Ic & faint$^b$ \\
		& &&&&&& & &  &  &  & & CC SNe\\ 
		\midrule

		1.2 +   1.1 &  0.80 & $2350(70)$ & $1.78(0.05) $ &$2.80(0.07)$  & $0.06(0.03)   $ & $0.04(0.02) $ &  y  &  y  &  y  &  y  &  y  &  y  &  y  \\
		1.2 + 1.3 & 0.80 & $2400(70)$& $\cdots$ &$2.81(0.08)$ & $\cdots$ & $\cdots$   &  y  &  y  &  y  &  y  &  y  &  y  &  y \\ [2pt]
		2.1 + 2.2 & 0.73 & 143 (17) & 2.44(0.06) & 2.11(0.05) & 0.90(0.05) & 0.54(0.04) &     y  &  y  &  y  &  y  &  y  &  y  &  y  \\ 
		2.1 + 2.3 & 0.73 & 89(17)  & $\cdots$& 2.75(0.07)&  $\cdots$ & $\cdots$   &  y  &  y  &  y  &  y  &  y  &  y  &  y\\ 
		2.1 + 2.4 & 0.73 & 92(17)  & $\cdots$ & 4.11(0.10)&   $\cdots$ & $\cdots$   &  y  &  y  &  y  &  y  &  y  &  y  &  y\\ 
		2.1 + 2.5 & 0.73 & 61(17)  & $\cdots$  & 4.23(0.10)&  $\cdots$ & $\cdots$   &  y  &  y  &  y  &  y  &  y  &  y  &  y\\ [2pt]
		3.3 + 3.1 & 1.95 & 660(60) & 2.40(0.07) & 3.59(0.19) & 0.31(0.13) & 0.18(0.07) &  y  &  y  &  y  &  y  &  y  &  y  &  y\  \\ 
		3.3 + 3.2 & 1.95 & 910(100) &$\cdots$  &3.69(0.18)&   $\cdots$ & $\cdots$   &  y  &  y  &  y  &  y  &  y  &  y  &  y\\[2pt] 
		4.3 + 4.1 & 1.27 & 800(90) &1.45(0.03)&  1.62(0.05)& 0.64(0.19) & 0.36(0.11) &  y  &  y  &  y  &  y  &  y  &  y  &  y \\ 
		4.3 + 4.2 & 1.27 & 7080(70) & $\cdots$& 1.74(0.05)  &  $\cdots$ & $\cdots$   &  y  &  y  &  y  &  y  &  y  &  y  &  y\\[2pt] 
		5.4 + 5.1 & 1.28 & 94(4) & 4.21(0.13)& 2.99(0.08)& 0.12(0.02) & 0.07(0.01) & y  &  y  &  y  &  y  &  y  &  y  &  y  \\ 
		5.4 + 5.2 & 1.28 & 104(8)  & $\cdots$& 4.09(0.09)& $\cdots$ & $\cdots$   &  y  &  y  &  y  &  y  &  y  &  y  &  y\\ 
		5.4 + 5.3 & 1.28 & 135(7)  &  $\cdots$& 4.27(0.12)& $\cdots$ & $\cdots$   &  y  &  y  &  y  &  y  &  y  &  y  &  y\\[2pt]  
		6.3 + 6.1 & 1.06 & 8320(110)& 1.41(0.03)& 1.68(0.04)& 3.16(0.16) & 2.06(0.20) &  y  &  y  &  y  &  y  &  y  &  y  &  y   \\
		6.3 + 6.2 & 1.06 & 6720(100) & $\cdots$ & 1.71(0.04) &  $\cdots$ & $\cdots$   &  y  &  y  &  y  &  y  &  y  &  y  &  y \\[2pt]  
		*7.4 + 7.1 & 2.75 & 36520(300) & 1.29(0.04)& 2.79(0.17)& 4.70(1.5) & 2.7(0.8) & y  &  y  &  y  &  y  &  y  &  y  &    \\ 
		*7.4 + 7.2 & 2.75 & 36560(300) &  $\cdots$ & 2.80(0.16)& $\cdots$ & $\cdots$   &  y  &  y  &  y  &  y  &  y  &  y  &  \\ 
		*7.4 + 7.3 & 2.75 & 34650(300) & $\cdots$ & 1.14(0.04)&   $\cdots$ & $\cdots$   &  y  &  y  &  y  &  y  &  y  &  y  &  \\ 
		*7.4 + *7.5 & 2.75 & 31040(300)  & $\cdots$ & 1.57(0.04) &  $\cdots$ & $\cdots$   &  y  &  y  &  y  &  y  &  y  &  y  &  \\ [2pt]  

		9.3 + 9.1 & 1.52 & 10830(170) &1.52(0.03)& 1.48(0.03)& 0.11(0.03) & 0.06(0.01) &  y  &  y  &  y  &  y  &  y  &  y  &  y    \\ 
		9.3 + 9.2 & 1.52 & 15500(160)&   $\cdots$& 1.61(0.03)&  $\cdots$ & $\cdots$   &  y  &  y  &  y  &  y  &  y  &  y  &  y \\  [2pt]

		*14.5 + 14.1 & 3.13 & 32930(170) & 0.89(0.03) &2.37(0.06) & 0.33(0.11) & 0.19(0.06) &  y  &    &  y  &  y  &  y  &    &   \\ 
		*14.5 + 14.2 & 3.13 & 33010(190) & $\cdots$ &  0.88(0.07)& $\cdots$ & $\cdots$   &  y  &    &  y  &  y  &  y  &    &   \\ 
		*14.5 + 14.3 & 3.13 & 30320(180) & $\cdots$ &1.87(0.04) & $\cdots$ & $\cdots$   &  y  &    &  y  &  y  &  y  &   &   \\ 
		*14.5 + *14.4 & 3.13 & 25800(170) &$\cdots$  & 1.96(0.06) &$\cdots$ & $\cdots$   &  y  &    &  y  &  y  &  y  &   &   \\[2pt]    
		*15.5 + 15.1 & 3.71 & 23090(150) & 1.06(0.02) &0.97(0.02) &  0.23(0.04) & 0.13(0.02) &y  &    &  y  &  y  &   &   &    \\ 
		*15.5 + 15.2 & 3.71 & 18990(130) &  $\cdots$& 1.34(0.04) &$\cdots$ & $\cdots$   &  y  &    &  y  &  y  &    &    &   \\ 
		*15.5 + 15.3 & 3.71 & 22950(150)  &   $\cdots$& 1.97(0.06) & $\cdots$ & $\cdots$   &  y  &    &  y  &  y  &    &    &  \\ 
		*15.5 + *15.4 & 3.71 & 4980(250)  &   $\cdots$ & 1.61(0.04)& $\cdots$ & $\cdots$   &  y  &    &  y  &  y  &   &   &  \\ [2pt]   	
		*16.3 + 16.1 & 3.77 & 13160(200) &  1.50(0.03) & 2.63(0.07) & 0.07(0.04) & 0.04(0.02) & y  &    &  y  &  y  &   &    &  \\ [2pt]   

		*21.3 + 21.1 & 1.26 & 12230(130) & 1.27(0.03) &  3.73(0.09) &0.24(0.13) & 0.14(0.08) &  y  &  y  &  y  &  y  &  y  &  y  &  y   \\ 
		*21.3 + 21.2 & 1.26 & 12300(130) & $\cdots$ & 3.89(0.10) &  $\cdots$ & $\cdots$   &  y  &  y  &  y  &  y  &  y  &  y  &  y \\ [2pt]   	
		*22.3 + 22.1 & 3.13 & 26940(210) & 0.91(0.03) & 2.25(0.06)& 0.15(0.08) & 0.08(0.05) &  y  &    &  y  &  y  &  y  &    &  \\ 
		*22.3 + *22.2 & 3.13 & 2260(140) & $\cdots$ & 2.19(0.06) &$\cdots$ & $\cdots$   &  y  &    &  y  &  y  &  y  &    &   \\  
		
		\bottomrule
	\end{tabular} 
	\tablefoot{	
	$^{\rm a}$ Images labeled with an asterisk (*) do not have spectroscopic redshifts. \\[2pt]
	$^{\rm b}$ Population of intrinsically faint CC SNe ($\rm M_B >-15$) which could be contributing importantly in the CC SN rates. See \eg \citet{2015ApJ...813...93S} for a discussion.}
}
\end{table*}

\subsection{Search for supernovae in the multiply-imaged galaxies behind A370 in the HFF data}

In addition to the HAWK-I data, we also looked for SNe in the HFF imaging data. The majority of the A370 HFF data were observed during 2015 and 2016, using both the HST/ACS and HST/WFC3 cameras \citep{2017ApJ...837...97L}.  These data are supplemented by additional, older archival images, as described in L17.  The observations are listed in Table~\ref{HFF_data} and~\ref{HFF_data1} along with their exposure times.  While the HFF team already has performed a live search for transients, these efforts have not resulted in the discovery of any strongly lensed SN in the A370 data\footnote{Steven Rodney, private communication.}.
However, they have reported the discovery of one \snia spectroscopically classified in one of the cluster member galaxies at $z=0.375$ \citep{2016ATel.8545....1G}. 

 In order to verify that there are no strongly lensed SNe in the HST A370 data, we downloaded the public images and recombined them to perform an independent transient search.
 The images were reduced via the standard Pyraf/STSDAS pipeline. Epochs that were closer than 14 days in time were aligned and combined to obtain better image depths using standard DrizzlePac\footnote{http://drizzlepac.stsci.edu/} routines, supplemented by SCAMP \citep{2006ASPC..351..112B} and SWarp \citep{2002ASPC..281..228B} to improve the astrometric solution and image stacking process.  The observations and image depths are compared with those of the HAWK-I survey in Figure~\ref{maglimit} (see also Tables~\ref{HFF_data} and \ref{HFF_data1}). After combining the data, we visually inspected the images for transients and found none, thus confirming the result of the HFF team. In Appendix~\ref{sec:patches}, we show examples of  image stamps of the HST data zoomed-in on the multiply-imaged galaxies together with those of the HAWK-I data. To follow the HAWK-I analysis as closely as possible, we take the image depth of each combined stack to be the $5\sigma$ point-source limiting magnitude.  We calculate this value by averaging the flux in a series of 0.4$\arcsec$~ circular apertures placed in so-called blank areas of sky throughout the field of view, normalised to the HFF magnitude zero-points for each band.  To be consistent from stack to stack, we keep the locations of the apertures fixed.  While we try to include all apertures in every stack, there are a few instances (due to differences in pointing solutions or the field-of-view size between the ACS and WFC3) where some apertures fall either completely or partially outside of the image exposure area.  In those cases, we simply remove the affected apertures from the average and proceed to calculate the image depth exactly as before.

 Following the same procedure as for the HAWK-I survey, we also calculate the expectation of observing a lensed SN in the multiply-imaged galaxies behind A370 in the HFF data. As previously, we focus on the galaxy systems that have secure redshifts. For SNe~Ia and SNe~IIP, the HFF A370 campaign was sensitive to 27 galaxies originating from 8 unique systems (1, 2, 3, 4, 5, 6, 9, and 21) with redshifts ranging between $0.72<z<1.95$. Thanks to the better depth, the volume where the HFF could detect CC SNe is significantly larger. Few examples of simulated light curves are shown in Figure~\ref{maglimit}. The result is  $N_{\rm CC}=0.39\pm0.09$ and   $N_{\rm Ia}=0.04 \pm0.01$ for CC~SNe and  \sneia, respectively.  The improvement in the image depth does not  improve greatly the expected number of SNe~Ia compared to the HAWK-I survey, mainly because both surveys were sensitive to similar number of galaxies. The HFF A370 observations were made over two years in which the epochs in different bands are close in time (see Figure~\ref{maglimit}). Therefore, from this aspect the HFF survey does not have significantly improved control time for SNe~Ia, thus the expected number of lensed SNe Ia remains largely unchanged. 

We can extend this estimate to the remaining HFF clusters,  Abell~2744, MACSJ0416.1-2403, MACSJ0717.5+3745, MACSJ1149.5+2223, and Abell~S1063. In the field of view of each of these galaxy clusters there are between 9 and 51 unique systems with 30 to 165 multiple images \citep{2011MNRAS.410.1939Z, 2015MNRAS.446.4132J,2015A&A...574A..11K,2016A&A...588A..99L, 2018MNRAS.473..663M}. For the purpose of making a rough estimate of the  expectations of the entire HFF survey for lensed SNe, we can assume that the other five galaxy clusters are represented by A370 properties. That means that the total number of expected lensed SNe was $\sim2.6$, in agreement with the observation of SN~Refsdal, once the statistical uncertainty is taken into account.

\section{Expectations for the future}
Several space-based telescopes are being planned such as the Wide-Field Infrared Survey Telescope (WFIRST; \citealt{2015arXiv150303757S}) and James Webb Space Telescope (JWST; \citealt{2006SSRv..123..485G}). With a 6.5 m aperture and an extensive infrared filter set, JWST, scheduled for launch in spring 2019, will have unprecedented resolution and sensitivity.  Since the NIRCam instrument aboard the JWST will be aimed, among others, at discovering SNe from the earliest stages of the universe (with the help of gravitational telescopes),  it is interesting to study what the prospects will be for multiply-imaged SNe. The WFIRST Wide-field instrument (WFI) will also have similar NIR filters, though at rather different effective wavelengths\footnote{\url{https://wfirst.ipac.caltech.edu/sims/Param_db.html}}. We focus on two JWST/NIRCam filters, F115W and F150W. For comparison, the effective wavelength of the JWST/NIRCam F115W lies between the WFIRST/WFI Y106 and J129, while the JWST/NIRCam F150W is similar to WFIRST/WFI H158 filter. 

\begin{figure}[htbp]
	\begin{center}
		\includegraphics[width=\columnwidth]{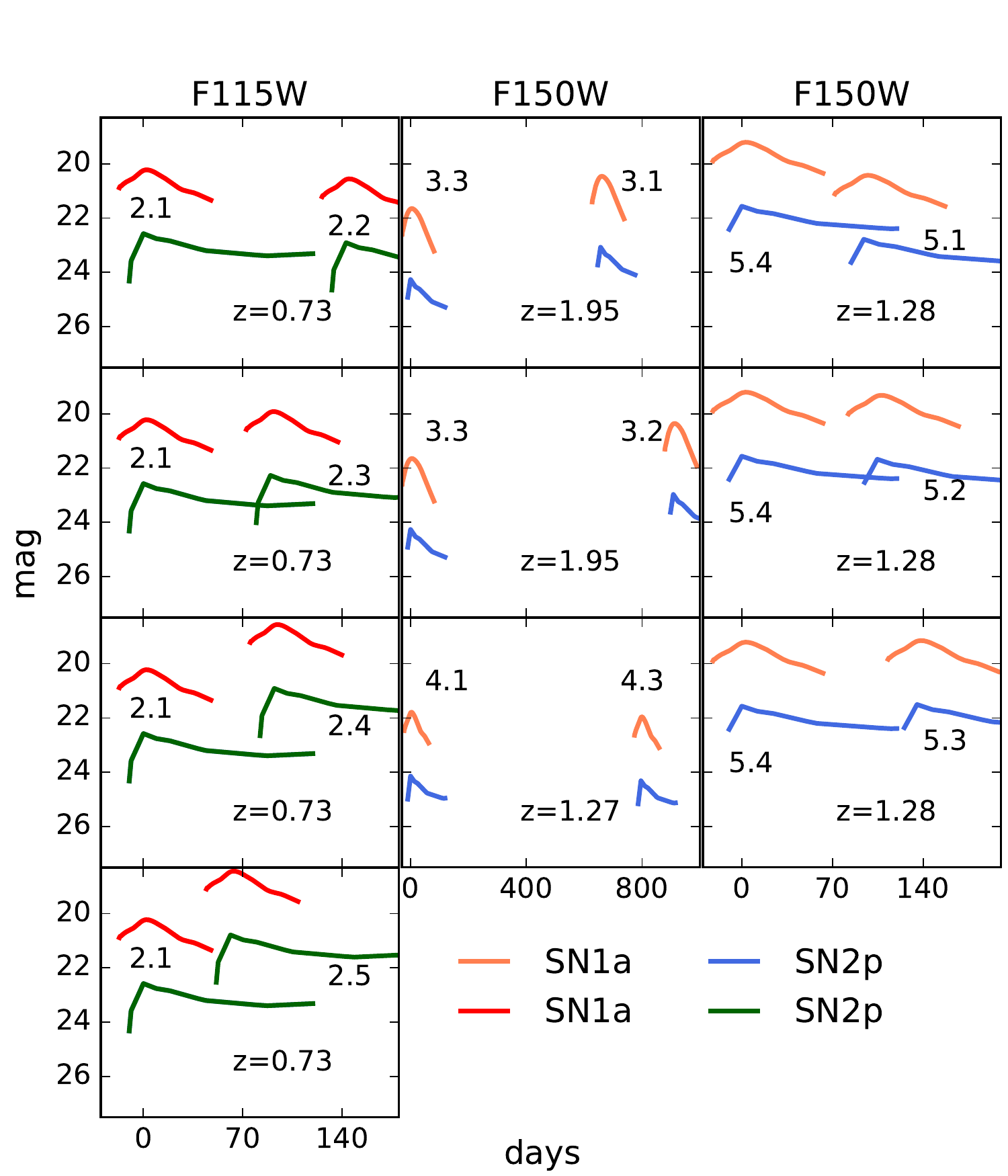}
		\caption{Examples of simulated light curves of SNe Ia and IIP for pairs of images of strongly lensed galaxies behind A370 (see Table~\ref{table:multi}). The first four images are in F115W band, while the rest is in F150W, similar to the filter bands planned for the JWST/NIRCam instrument. If a SN explodes in one of the galaxies, the explosion should appear in the corresponding images, thus allowing for scheduling observations accordingly. }
		\label{times}
	\end{center}
\end{figure}

First, we simulate the expected brightness of SNe exploding in multiply-imaged galaxies behind A370. In Figure~\ref{times} we show examples of synthetic light curves of SNe Type Ia and IIP in two bands (F115W and F150W). From the figure, we see that typical SNe of both types should be easily detected, as the JWST/NIRCam instrument is expected to reach a  depth of 27.5 mag (5$\sigma$) in both F115W and F150W, in just 1 hour of exposure time \citep{2012SPIE.8442E..2NB}. 
	
Next, we simulate a feasible monitoring survey targeting A370 using the NIRCam F115W and F150W filters. We repeat the simulation for a survey targeting another galaxy cluster, A1689, which was the subject of our work in P16.  As in the previous section, we only consider the systems with secure redshifts.
As more data are being collected, new multiple image systems might be detected. Therefore, the predictions presented below are to be considered as lower limits. There are 69 images originating from 19 unique galaxies that fulfill this criteria behind A1689 (see table 3 in \citealt{Third}) and 47 images from 13 unique galaxies behind A370 (Table~\ref{table:multi}). To calculate the expectations of the galaxies behind A1689, we used the SN rates presented in \citet{Third}. We set the duration of the survey to be one year and assume either 4 or 12 visits to each cluster per year. Using the procedure highlighted in the previous section, we then calculate the total expected number of SNe seen in both clusters over the duration of the monitoring campaign.  These results are shown in Table~\ref{table:JWST}.

We can now compare the gravitational telescope properties of A370 with those of A1689, which could be useful when deciding which clusters will be targeted with JWST for a dedicated multi-year search with its relatively small field of view of $2.2\arcmin \times 2.2\arcmin$.  Based on our analysis, we expect that the observable number of SNe exploding in the multiply-imaged galaxies behind A370 will be lower, compared to A1689 (see Table~\ref{table:JWST}). There are several factors that influence this outcome. One of these is the number and the redshift of the multiply-imaged galaxies. A370 has fewer observable multiple-image galaxies, 47 compared to 66 of A1689.
Another important factor is the magnification from the galaxy cluster lens. The median magnification at the positions of the multiply-lensed galaxies, are $\sim2.0$ mag, while for A1689 is almost one magnitude greater, $\sim2.9$. This is also the reason why
our NIR survey of A370 was less sensitive compared to that of A1689 in P16, i.e. the A1689 survey was sensitive to SNe exploding in galaxies at higher redshifts, compared to the survey targeting A370 (see also fig. 8 in P16).  This study is useful to optimise the search strategy, dedicating more epochs to one cluster or observing more clusters with fewer epochs. The results in Table \ref{table:JWST} convey that the most profitable strategy is to target more clusters that have many multiply-imaged galaxies such as A1689 and A370, rather than focussing more time on a single galaxy cluster.

\begin{table}
\caption{Expectations for lensed SNe in the multiply-imaged galaxies behind A1689 and A370 observed with JWST/NIRCam-like filters in one year.  The depth of the survey in both filters is assumed to be 27.5 mag, which is the limiting $5\sigma$ depth for a 1 hour exposure \citep{2012SPIE.8442E..2NB}. The separation between two epochs is set to 30 days. }\label{table:JWST} \centering \begin{tabular}{lccccc} \toprule  Filter   & N$_{epochs}$& N$_{\rm CC}^a$ & N$_{\rm Ia}^a$ & $z_{max}^b$ & N$_{gal}^c$\\
		
		  &/yr & && & \\
		\midrule
  & &	A370  && & \\
		\midrule

 F115W &4   & $0.11 \pm0.04$ & $0.02 \pm0.01$ & 3.77 & 47 \\
 F115W  &12   & $0.4\pm0.1 $ &  $0.06 \pm 0.03$ & 3.77 &  47 \\
F150W  &4   & $0.3\pm0.1$ & $0.02 \pm 0.01$ &3.77 & 47  \\
 F150W&12   & $0.9 \pm0.5$ & $0.06 \pm0.03$ & 3.77 & 47 \\

		\midrule
  & &		A1689  && & \\
		\midrule
F115W  &4  & $0.7 \pm 0.3$ & $0.13\pm 0.06$ & 3.05  &66 \\ 
F115W &12   & $1.0\pm0.5$ & $ 0.17 \pm 0.08$ & 3.05  &66 \\
F150W &4  & $1.0\pm0.5 $ & $ 0.14\pm 0.07$ & 3.05 & 66 \\
F150W &12  & $1.4\pm0.6$ & $0.17 \pm 0.08$ & 3.05 &66 \\

		\bottomrule
	\end{tabular} 
	\tablefoot{The errors in the N$_{\rm CC}$ and N$_{\rm Ia}$ originate from the propagated uncertainty in the SFR. \\
		$^a$ The number of expected \sne in the background galaxies with resolved multiple images and secure redshifts (see Table\ref{table:multi}). \\
		$^b$ The maximum redshift of the expected \sne. \\
		$^c$ Number of galaxies that can host observable \sne. \\
	}
\end{table}

\section{Summary and conclusions}
In this work, we have used the galaxy cluster A370 as a gravitational telescope to search for lensed SNe. Recently, a study improved the lensing model of A370 and secured redshifts of even more multiply-imaged galaxies that lie behind the cluster (L17). 
During 2012--2013, we obtained 16 epochs of ground-based NIR data of A370 with the aim of searching for SNe in these resolved multiply-imaged galaxies. This search did not yield in 
 SN discoveries, while the number of expected SN in these galaxies was $\sim 0.1$, summed for CC~SNe and  \sneia. 

We used multi-band HST photometry to infer the expectation of SN rates in the multiply-imaged galaxies. From the lensing model in L17, we computed the magnifications and time delays of the arrival of the images of these systems. This will be very useful for future work using new generation telescopes, such as space-based JWST and WFIRST. We simulated the light curves of SNe Ia and IIP that could explode in the multiply-imaged galaxies behind A370, and concluded that they will be detectable by the JWST/NIRCam in F115W and F150W. We also simulated possible JWST surveys to obtain the expected number of SNe that have multiple images behind both A1689 and A370, and concluded that A1689 offers better prospects. We also found that the strategy to spread the time over several clusters is more productive than investing more epochs targeting single galaxy cluster.

\begin{acknowledgements}
	We thank Steven Rodney and the FrontierSN team for providing information regarding their transient search in the HFF data. We thank M. Kriek for providing information regarding the FAST code. We thank Brandon Anderson for proofreading the manuscript.	The work is based, in part, on observations obtained at the ESO Paranal Observatory. RA and AG acknowledge support from the Swedish Research Council and the Swedish Space Board. The Oskar Klein Centre is funded by the Swedish Research Council. DJL acknowledges support from the European Research Council (ERC) starting grant.
	336736-CALENDS.
\end{acknowledgements}

\newpage
\appendix

\section{ACS/HST and WFC3/HST data of A370}\label{sec:appendix}

	\begin{table}[htb]

		\caption{Photometry from the ACS data in two optical bands, F435W and F814W, is listed here. There are observations with the F606W band, which are combined into only one epoch, so they were not considered when calculating the expectations. The line indicates that those observations were combined for a deeper image depth.}
		\label{HFF_data}
		\begin{tabular}{ccc}
			\toprule
				Date &\multicolumn{1}{c}{Exp. time} & $m_{lim}^a$ \\ 
			\midrule
			& \large{F435W} &  \\ 
			\midrule
			2015-12-19 & 5140 & ref\\ 
			2015-12-20 & 5140 & \\ 
			2015-12-21 & 5140 & \\ 
			2015-12-24 & 5140 & \\ 
			2015-12-26 & 5140 & \\ 
			2015-12-26 & 5140 & \\ 
			2015-12-27 & 5140 & \\ 
			2015-12-29 & 5140 & \\ 
			2016-01-02 & 5140 & \\ 
			\midrule
			2016-02-17 & 5140 & 27.15\\ 
			\midrule
			& \large{F814W} &  \\ 
			\midrule
			2009-07-16 & 3840 &ref \\ 
			\midrule
			2010-12-20 & 4720 & 27.31\\ 
			2010-12-20 & 4880 & \\ 
			\midrule
			2015-12-19 & 5080 & 27.98 \\ 
			2015-12-20 & 5080 & \\ 
			2015-12-21 & 5080 & \\ 
			2015-12-22 & 5146 & \\ 
			2015-12-24 & 5146 & \\ 
			2015-12-24 & 5080 & \\ 
			2015-12-26 & 5080 & \\ 
			2015-12-26 & 5080 & \\ 
			2015-12-27 & 5080 & \\ 
			2015-12-28 & 5146 & \\ 
			2015-12-29 & 5080 & \\ 
			2016-01-02 & 5080 & \\ 
			\midrule
			2016-01-03 & 5146 & 27.43\\ 
			2016-01-10 & 5146 & \\ 
			2016-01-11 & 5146 & \\ 
			2016-01-13 & 4981 & \\ 
			2016-01-17 & 4981 & \\ 
			\midrule
			2016-01-18 & 5146 & 27.39\\ 
			2016-01-20 & 5146 & \\ 
			2016-01-23 & 4981 & \\ 
			2016-01-28 & 5146 & \\ 
			\midrule
			2016-02-17 & 5146 & 27.00\\ 
			2016-02-17 & 5080 & \\ 
			\bottomrule

\end{tabular}
\end{table}

\begin{table}[htb]

	\caption{Photometry from the Wide-Field Camera 3 (WFC3) in four NIR bands: F105W, F125W, F140W, and F160W, is listed here. The line indicates that those observations were combined for a deeper image depth.}
	\label{HFF_data1}
	\begin{tabular}{ccc}
		\toprule
		Date &\multicolumn{1}{c}{Exp. time} & $m_{lim}^a$ \\ 
		\midrule
				& \large{F105W} &  \\ 
		\midrule
		2014-10-27 & 356 & ref\\ 
		2014-10-29 & 812 & \\ 
		\midrule
		2014-12-10 & 356 & 25.64\\ 
		2014-12-12 & 812 & \\ 
		\midrule
		2016-07-27 & 5612 &28.01 \\ 
		2016-07-28 & 5612 & \\ 
		2016-07-29 & 5612 & \\ 
		2016-08-02 & 5612 & \\ 
		2016-08-03 & 5612 & \\ 
		2016-08-04 & 5612 & \\ 
		2016-08-05 & 5612 & \\ 
		2016-08-07 & 5612 & \\
		\midrule  
		2016-08-21 & 5612 & 27.553\\ 
		\midrule 
		2016-09-03 & 5612 & 27.17\\ 
		2016-09-04 & 5612 & \\ 
		2016-09-06 & 5512 & \\ 
		\midrule
		& \large{F125W} &  \\ 
		\midrule
		2016-07-29 & 5512 & ref\\ 
		2016-08-02 & 5512 & \\ 
		2016-08-05 & 5512 & \\ 
		\midrule
		2016-09-08 & 5412 & 27.33\\ 
		2016-09-11 & 5412 & \\ 
		2016-09-11 & 5412 & \\ 
		\midrule

		& \large{F140W} &  \\ 
		\midrule
		2009-09-28 & 4235 & ref \\ 
		\midrule
		2014-10-27 & 812 & 25.70\\ 
		2014-10-29 & 762 & \\
		\midrule 
		2014-12-10 & 812 & 25.37\\ 
		2014-12-12 & 762 & \\ 
		\midrule
		2015-07-19 & 5512 & 27.04\\ 
		2016-07-28 & 5512 & \\ 
		2016-07-30 & 5512 & \\ 
		\midrule
		2016-09-09 & 5412 & 27.11\\ 
		2016-09-10 & 5412 & \\ 
		\midrule
		& \large{F160W} &  \\ 
		\midrule
		2010-12-19 & 2412 & ref\\ 
		\midrule
		2016-02-08 & 2412 & 26.06 \\
		2016-02-14 & 2412 & \\ 
		2016-02-22 & 2412 & \\ 
		\midrule
		2016-07-27 & 5512 & 27.11 \\ 
		2016-07-28 & 5512 & \\ 
		2016-07-29 & 5512 & \\ 
		2016-08-02 & 5512 & \\ 
		2016-08-03 & 5512 & \\ 
		2016-08-04 & 5512 & \\ 
		2016-08-05 & 5512 & \\ 
		2016-08-07 & 5512 & \\ 
		\midrule
		2016-08-21 & 5512 & 26.74\\ 
		2016-09-03 & 5512 & \\ 
		2016-09-04 & 5512 & \\ 
		\midrule
		2016-09-06 & 5512 & 26.23\\ 
			\bottomrule

\end{tabular}
\end{table}

\begin{sidewaystable*}
	\centering
	\caption{ACS/HST and WFC3/HST photometry of the multiply-imaged galaxies. }
	\label{photometry}

	\begin{tabular}{@{}lllllllllllllll@{}}
			\toprule
		ID   &       mag435  &err435  &mag606 & err606 & mag814&  err814 & mag105 & err105 & mag125  &err125  &mag140  &err140  &mag160 & err160 \\ \midrule
1.1 & 24.399 & 0.023 & 23.875 & 0.016 & 22.748 & 0.005 & 22.464 & 0.004 & 22.164 & 0.006 & 22.021 & 0.006 & 21.824 & 0.005  \\ 
1.2 & 25.292 & 0.023 & 24.811 & 0.017 & 23.73 & 0.006 & 23.503 & 0.006 & 23.228 & 0.009 & 23.102 & 0.008 & 22.914 & 0.008  \\ 
1.3 & 24.333 & 0.021 & 23.795 & 0.015 & 22.646 & 0.005 & 22.37 & 0.004 & 22.057 & 0.005 & 21.928 & 0.005 & 21.732 & 0.005  \\ 
2.1 & 23.391 & 0.014 & 22.693 & 0.009 & 21.657 & 0.003 & 21.037 & 0.002 & 20.522 & 0.003 & 20.228 & 0.002 & 19.873 & 0.002  \\ 
2.2 & 25.783 & 0.045 & 24.594 & 0.017 & 23.321 & 0.005 & 22.794 & 0.004 & 22.332 & 0.005 & 22.05 & 0.004 & 21.721 & 0.003  \\ 
2.3 & 24.414 & 0.025 & 23.397 & 0.012 & 21.886 & 0.003 & 21.165 & 0.002 & 20.612 & 0.002 & 20.282 & 0.002 & 19.925 & 0.001  \\ 
2.4 & 24.341 & 0.029 & 23.165 & 0.011 & 21.42 & 0.002 & 20.588 & 0.001 & 20.008 & 0.001 & 19.677 & 0.001 & 19.324 & 0.001  \\ 
2.5 & 24.341 & 0.029 & 23.165 & 0.011 & 21.42 & 0.002 & 20.588 & 0.001 & 20.008 & 0.001 & 19.677 & 0.001 & 19.324 & 0.001  \\ 
3.1 & 24.29 & 0.023 & 24.066 & 0.022 & 23.84 & 0.016 & 23.466 & 0.013 & 23.056 & 0.016 & 22.615 & 0.011 & 22.337 & 0.01  \\ 
3.2 & 24.059 & 0.016 & 23.878 & 0.017 & 23.61 & 0.012 & 23.222 & 0.011 & 22.793 & 0.014 & 22.421 & 0.01 & 22.151 & 0.01  \\ 
3.3 & 25.578 & 0.038 & 25.333 & 0.037 & 24.961 & 0.025 & 24.342 & 0.019 & 23.92 & 0.024 & 23.864 & 0.024 & 23.727 & 0.025  \\ 
4.1 & 24.919 & 0.02 & 24.679 & 0.019 & 24.104 & 0.01 & 23.539 & 0.008 & 23.279 & 0.012 & 23.169 & 0.011 & 22.925 & 0.01  \\ 
4.2 & 24.79 & 0.016 & 24.622 & 0.016 & 24.151 & 0.009 & 23.604 & 0.008 & 23.368 & 0.012 & 23.271 & 0.011 & 23.001 & 0.01  \\ 
4.3 & 25.003 & 0.021 & 24.85 & 0.02 & 24.327 & 0.012 & 23.745 & 0.009 & 23.515 & 0.013 & 23.389 & 0.013 & 23.121 & 0.011  \\ 
5.1 & 26.252 & 0.06 & 25.875 & 0.049 & 25.049 & 0.021 & 24.136 & 0.012 & 23.818 & 0.017 & 23.563 & 0.015 & 23.318 & 0.013  \\ 
5.2 & 25.169 & 0.045 & 24.703 & 0.035 & 23.827 & 0.014 & 23.004 & 0.007 & 22.654 & 0.01 & 22.439 & 0.009 & 22.202 & 0.008  \\ 
5.3 & 25.325 & 0.043 & 24.917 & 0.035 & 24.058 & 0.014 & 23.269 & 0.009 & 22.951 & 0.013 & 22.746 & 0.011 & 22.482 & 0.01  \\ 
5.4 & 25.325 & 0.043 & 24.917 & 0.035 & 24.058 & 0.014 & 23.269 & 0.009 & 22.951 & 0.013 & 22.746 & 0.011 & 22.482 & 0.01  \\ 
6.1 & 27.753 & 0.483 & 26.07 & 0.121 & 24.064 & 0.017 & 22.777 & 0.006 & 21.946 & 0.006 & 21.486 & 0.004 & 21.07 & 0.003  \\ 
6.2 & 26.992 & 0.268 & 25.81 & 0.096 & 23.609 & 0.012 & 22.361 & 0.005 & 21.568 & 0.004 & 21.12 & 0.003 & 20.722 & 0.002  \\ 
6.3 & 27.807 & 0.488 & 26.132 & 0.124 & 23.966 & 0.015 & 22.66 & 0.006 & 21.849 & 0.005 & 21.401 & 0.004 & 20.982 & 0.003  \\ 
7.1 & 24.968 & 0.043 & 23.897 & 0.019 & 23.339 & 0.009 & 23.247 & 0.009 & 22.914 & 0.012 & 22.542 & 0.009 & 22.243 & 0.007  \\ 
7.2 & 24.968 & 0.043 & 23.897 & 0.019 & 23.339 & 0.009 & 23.247 & 0.009 & 22.914 & 0.012 & 22.542 & 0.009 & 22.243 & 0.007  \\ 
7.3 & 26.485 & 0.07 & 25.548 & 0.034 & 24.789 & 0.015 & 24.573 & 0.017 & 24.217 & 0.023 & 23.894 & 0.018 & 23.592 & 0.015  \\ 
7.4 & 26.779 & 0.084 & 25.584 & 0.034 & 24.963 & 0.018 & 24.745 & 0.019 & 24.341 & 0.025 & 23.949 & 0.019 & 23.561 & 0.015  \\ 
7.5 & 26.223 & 0.066 & 25.066 & 0.026 & 24.487 & 0.014 & 24.283 & 0.015 & 23.88 & 0.019 & 23.465 & 0.014 & 23.091 & 0.011  \\ 
9.1 & 26.954 & 0.065 & 26.891 & 0.069 & 26.62 & 0.05 & 26.298 & 0.048 & 25.263 & 0.035 & 25.217 & 0.036 & 25.432 & 0.049  \\ 
9.2 & 27.235 & 0.079 & 27.136 & 0.086 & 26.792 & 0.057 & 26.473 & 0.057 & 25.5 & 0.042 & 25.509 & 0.046 & 25.648 & 0.059  \\ 
9.3 & 27.119 & 0.083 & 26.798 & 0.073 & 26.401 & 0.047 & 26.084 & 0.046 & 25.153 & 0.036 & 25.123 & 0.038 & 25.196 & 0.046  \\ 
14.1 & 26.707 & 0.08 & 25.663 & 0.036 & 25.34 & 0.024 & 25.391 & 0.034 & 25.166 & 0.05 & 25.018 & 0.046 & 24.742 & 0.04  \\ 
14.2 & 28.869 & 0.221 & 27.327 & 0.062 & 27.187 & 0.049 & 27.157 & 0.065 & 26.907 & 0.096 & 26.775 & 0.093 & 26.541 & 0.08  \\ 
14.3 & 27.219 & 0.114 & 26.089 & 0.049 & 25.769 & 0.033 & 25.688 & 0.045 & 25.479 & 0.065 & 25.286 & 0.062 & 24.998 & 0.051  \\ 
14.4 & 27.367 & 0.102 & 26.325 & 0.046 & 25.917 & 0.029 & 25.911 & 0.04 & 25.891 & 0.071 & 25.643 & 0.059 & 25.427 & 0.057  \\ 
14.5 & 28.017 & 0.163 & 26.812 & 0.061 & 26.724 & 0.052 & 26.726 & 0.069 & 26.568 & 0.112 & 26.21 & 0.087 & 26.164 & 0.093  \\ 
\bottomrule

	\end{tabular}
\end{sidewaystable*}
\clearpage
\newpage

\begin{sidewaystable*}
	\centering
	\caption{ACS/HST and WFC3/HST photometry of the multiply-imaged galaxies.}

	\begin{tabular}{lllllllllllllll }

		\toprule
		ID   &       mag435  &err435  &mag606 & err606 & mag814&  err814 & mag105 & err105 & mag125  &err125  &mag140  &err140  &mag160 & err160 \\  \midrule
		15.1 & 29.615 & 0.61 & 28.059 & 0.17 & 27.957 & 0.137 & 27.755 & 0.158 & 27.603 & 0.241 & 27.338 & 0.213 & 26.987 & 0.167  \\ 
		15.2 & 99.102 & 0.0 & 28.706 & 0.28 & 28.72 & 0.239 & 28.968 & 0.313 & 28.516 & 0.377 & 28.409 & 0.371 & 28.319 & 0.366  \\ 
		15.3 & 31.37 & 0.0 & 28.78 & 0.278 & 27.888 & 0.11 & 28.015 & 0.166 & 27.538 & 0.201 & 27.201 & 0.152 & 27.176 & 0.168  \\ 
		15.4 & 31.087 & 0.0 & 28.912 & 0.231 & 28.565 & 0.155 & 28.687 & 0.232 & 28.52 & 0.364 & 28.435 & 0.353 & 29.578 & 0.0  \\ 
		15.5 & 28.608 & 0.247 & 29.028 & 0.412 & 27.78 & 0.12 & 27.275 & 0.101 & 26.869 & 0.126 & 26.951 & 0.148 & 26.629 & 0.122  \\ 
		16.1 & 28.953 & 0.397 & 27.782 & 0.158 & 26.993 & 0.07 & 26.79 & 0.079 & 26.546 & 0.114 & 26.448 & 0.117 & 26.309 & 0.111  \\ 
		16.3 & 29.438 & 0.205 & 28.56 & 0.24 & 27.908 & 0.124 & 28.042 & 0.168 & 27.89 & 0.298 & 27.398 & 0.187 & 27.195 & 0.168  \\ 
		21.1 & 25.915 & 0.054 & 25.612 & 0.048 & 25.104 & 0.027 & 24.829 & 0.031 & 24.791 & 0.055 & 24.711 & 0.055 & 24.364 & 0.043  \\ 
		21.2 & 25.915 & 0.054 & 25.612 & 0.048 & 25.104 & 0.027 & 24.829 & 0.031 & 24.791 & 0.055 & 24.711 & 0.055 & 24.364 & 0.043  \\ 
		21.3 & 26.28 & 0.051 & 25.816 & 0.039 & 24.851 & 0.015 & 24.058 & 0.009 & 23.734 & 0.013 & 23.526 & 0.011 & 23.309 & 0.011  \\ 
		22.1 & 27.445 & 0.102 & 26.494 & 0.049 & 26.301 & 0.037 & 26.443 & 0.069 & 26.256 & 0.108 & 26.218 & 0.11 & 25.955 & 0.094  \\ 
		22.2 & 27.254 & 0.089 & 26.391 & 0.048 & 26.007 & 0.031 & 26.173 & 0.05 & 25.885 & 0.07 & 25.788 & 0.067 & 25.55 & 0.063  \\ 
		22.3 & 28.45 & 0.184 & 27.959 & 0.131 & 27.574 & 0.085 & 27.808 & 0.14 & 27.348 & 0.168 & 27.266 & 0.171 & 27.095 & 0.163  \\
		 \bottomrule
	\end{tabular}
\end{sidewaystable*}
\clearpage
\newpage
\clearpage
\newpage

\section{HAWKI-J subtractions and WFC3/HST data of A370}\label{sec:patches}

Examples of HAWK-I data used for the transient search together with the HST/WFC3 data. The first three columns show $8\times8 \arcsec$ image (new and reference) and subtraction stamps from the \Jband-band HAWK-I data. The forth column is the HST/WFC3 image in F125W band. The location of the multiply-imaged galaxies are indicated by the red markers.

\begin{figure}[htbp]
		\includegraphics[width=0.8\textwidth]{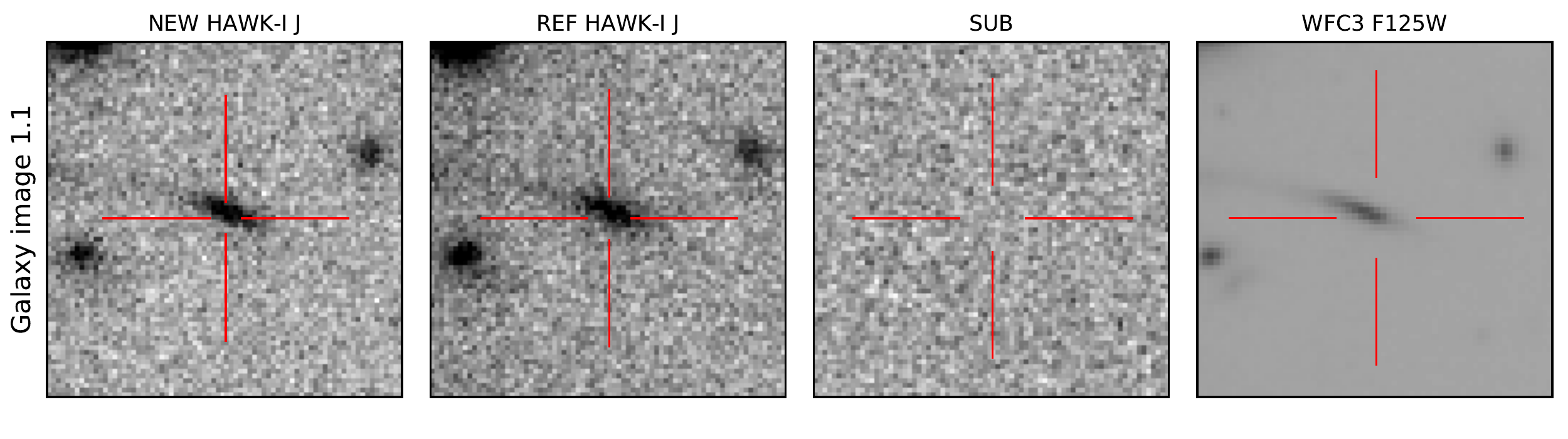}

		\label{f6.pdf}
\end{figure}
\begin{figure} [htbp]

		\includegraphics[width=0.8\textwidth]{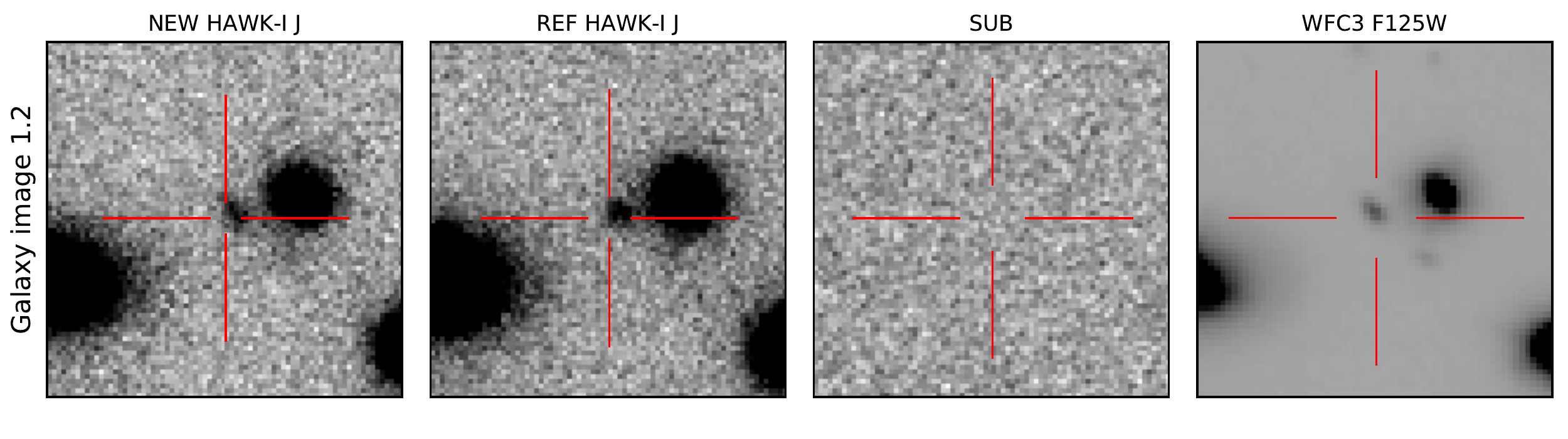}
		\label{f8.pdf}
\end{figure}
\begin{figure} [htbp]
		\includegraphics[width=0.8\textwidth]{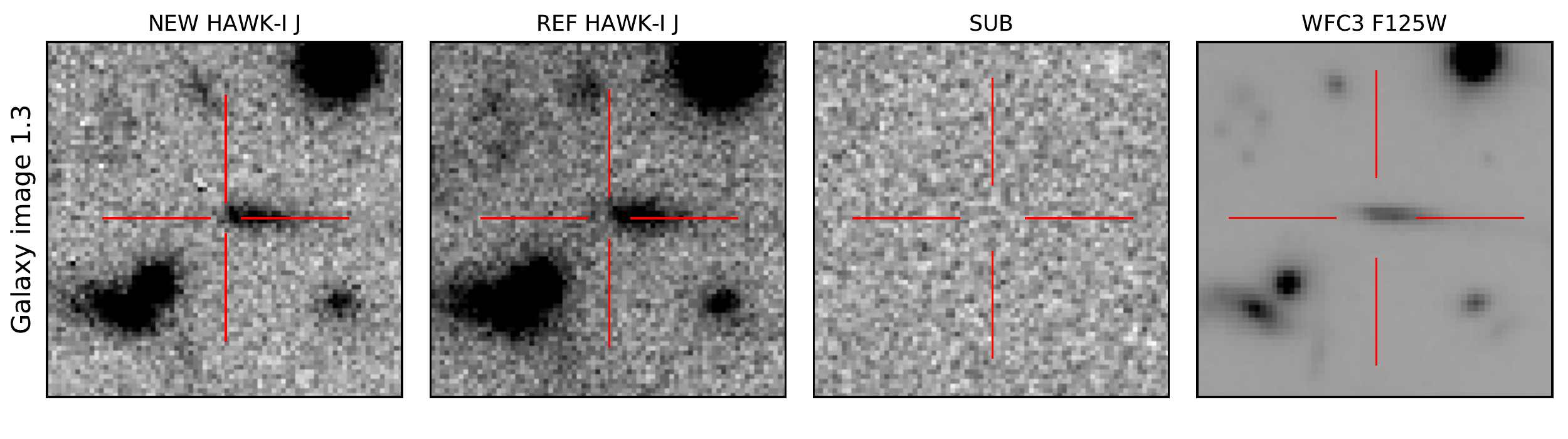}
		\label{f10.pdf}
\end{figure}
\begin{figure} [htbp]
		\includegraphics[width=0.8\textwidth]{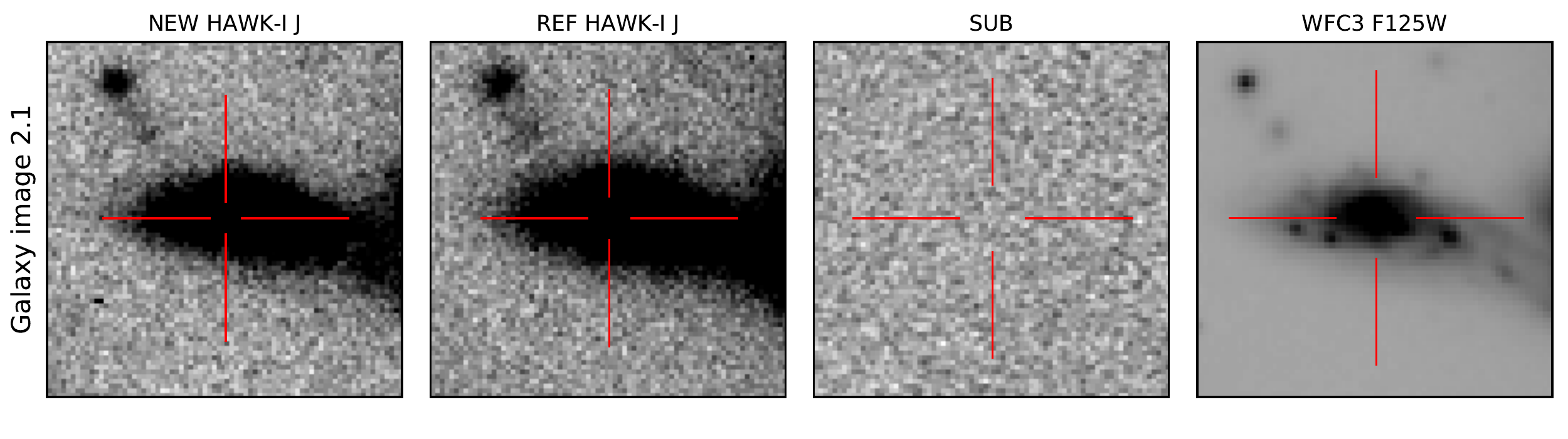}
		\label{f12.pdf}
\end{figure}
\begin{figure*} [htbp]
	\begin{center}
		\includegraphics[width=0.8\textwidth]{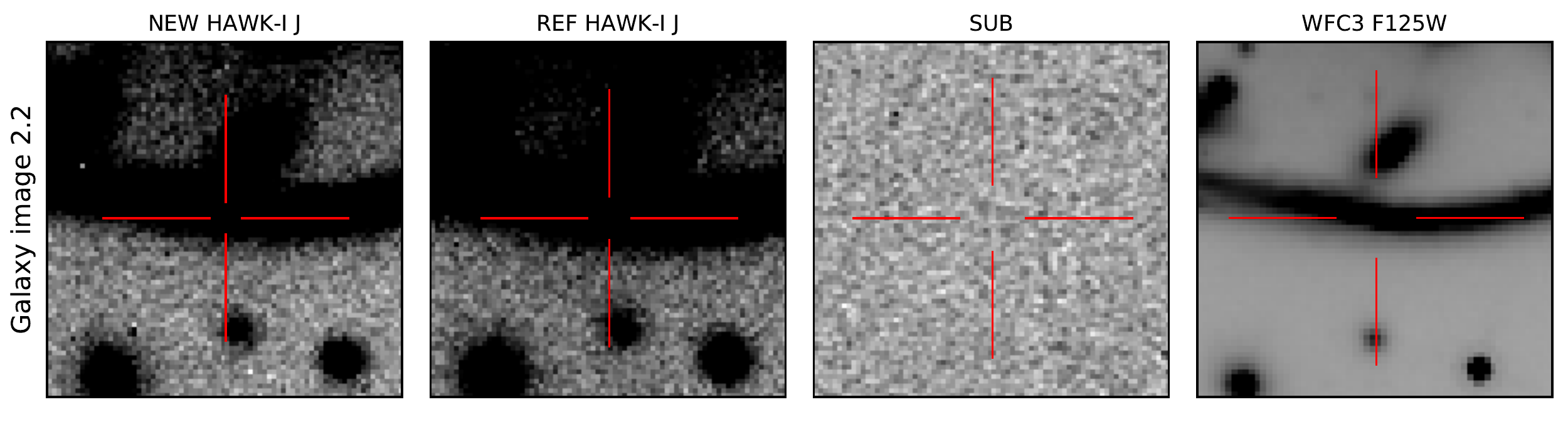}
		\label{f14.pdf}
	\end{center}
\end{figure*}
\begin{figure*} [htbp]
	\begin{center}
		\includegraphics[width=0.8\textwidth]{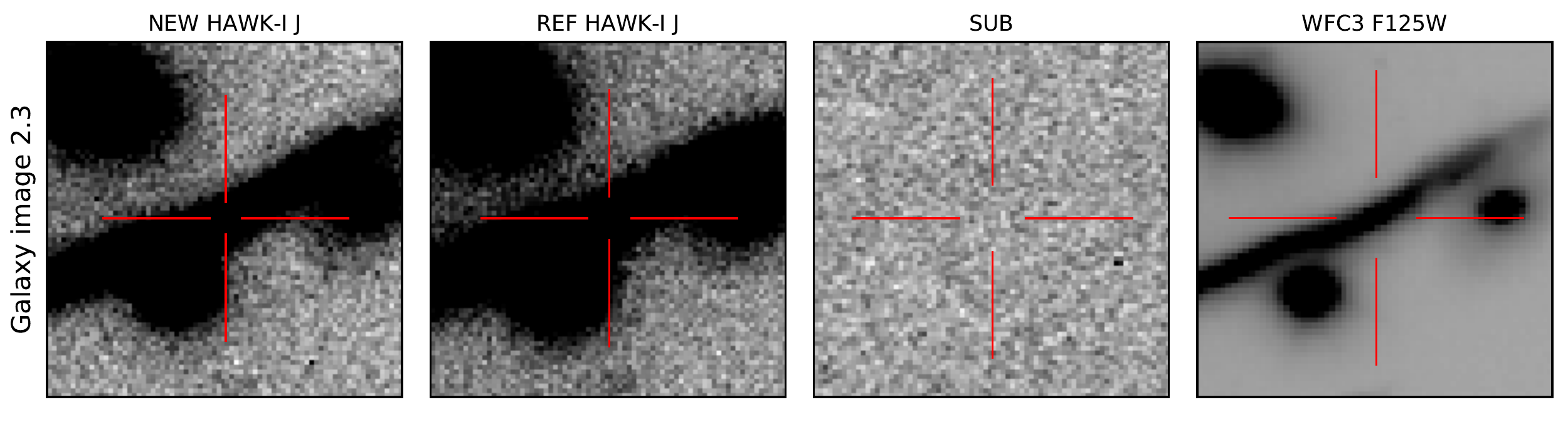}
		\label{f16.pdf}
	\end{center}
\end{figure*}
\begin{figure*} [htbp]
	\begin{center}
		\includegraphics[width=0.8\textwidth]{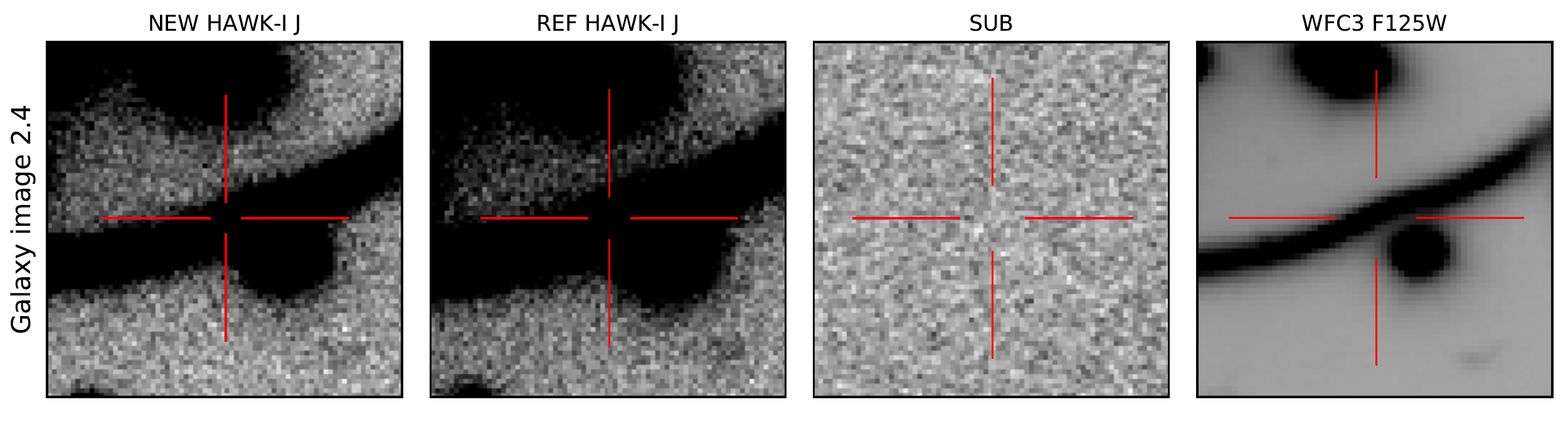}
		\label{f18.pdf}
	\end{center}
\end{figure*}
\begin{figure*} [htbp]
	\begin{center}
		\includegraphics[width=0.8\textwidth]{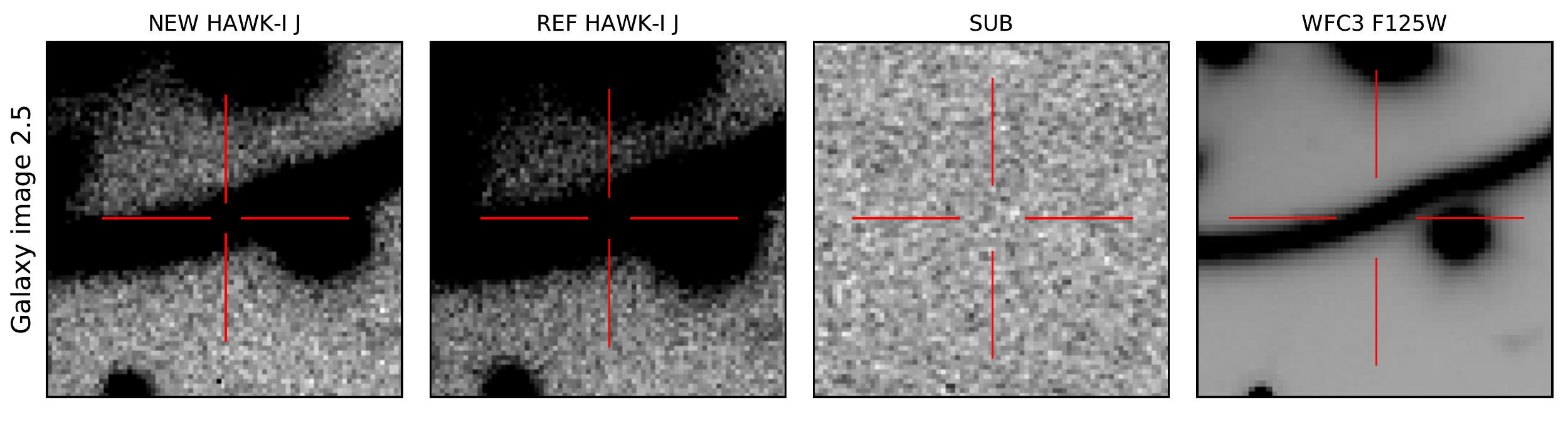}
		\label{f20.pdf}
	\end{center}
\end{figure*}
\begin{figure*} [htbp]
	\begin{center}
		\includegraphics[width=0.8\textwidth]{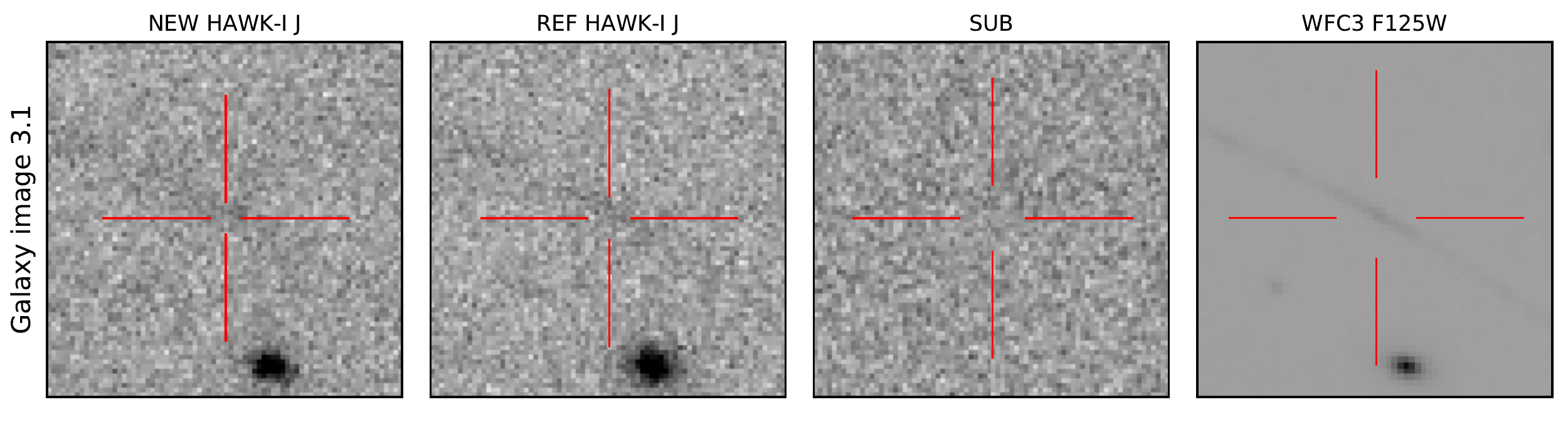}
		\label{f22.pdf}
	\end{center}
\end{figure*}
\begin{figure*} [htbp]
	\begin{center}
		\includegraphics[width=0.8\textwidth]{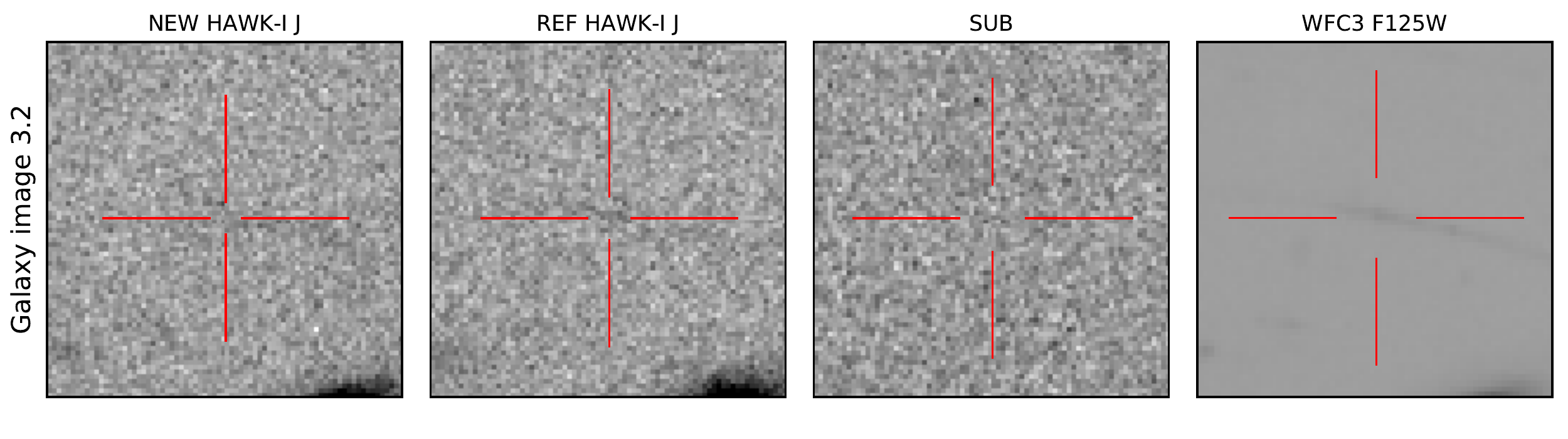}
		\label{f24.pdf}
	\end{center}
\end{figure*}
\begin{figure*} [htbp]
	\begin{center}
		\includegraphics[width=0.8\textwidth]{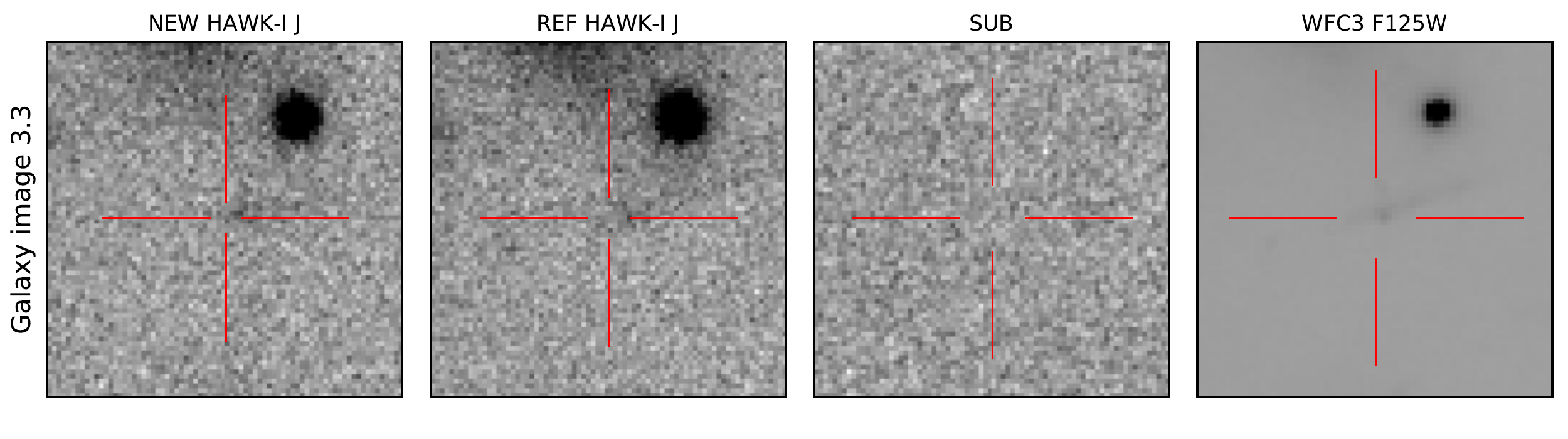}
		\label{f26.pdf}
	\end{center}
\end{figure*}
\begin{figure*} [htbp]
	\begin{center}
		\includegraphics[width=0.8\textwidth]{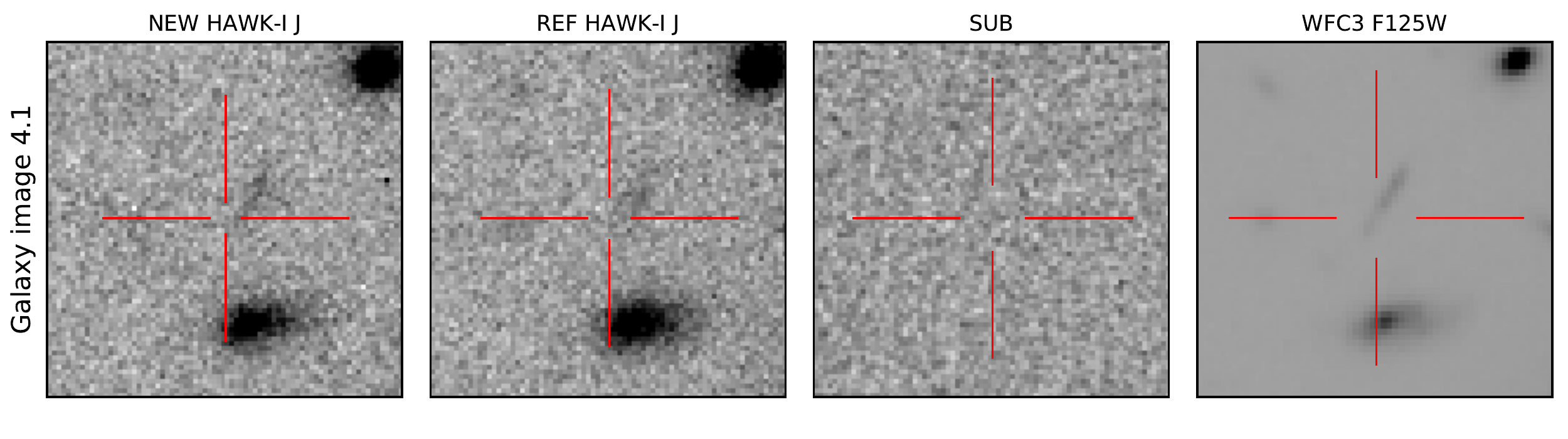}
		\label{f28.pdf}
	\end{center}
\end{figure*}
\begin{figure*} [htbp]
	\begin{center}
		\includegraphics[width=0.8\textwidth]{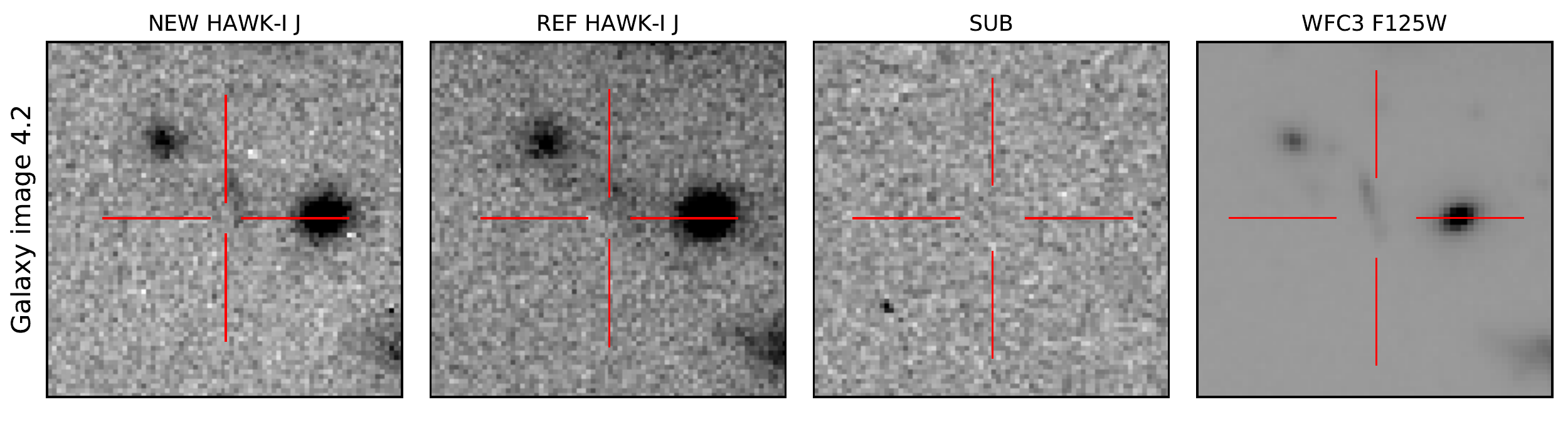}
		\label{f30.pdf}
	\end{center}
\end{figure*}
\begin{figure*} [htbp]
	\begin{center}
		\includegraphics[width=0.8\textwidth]{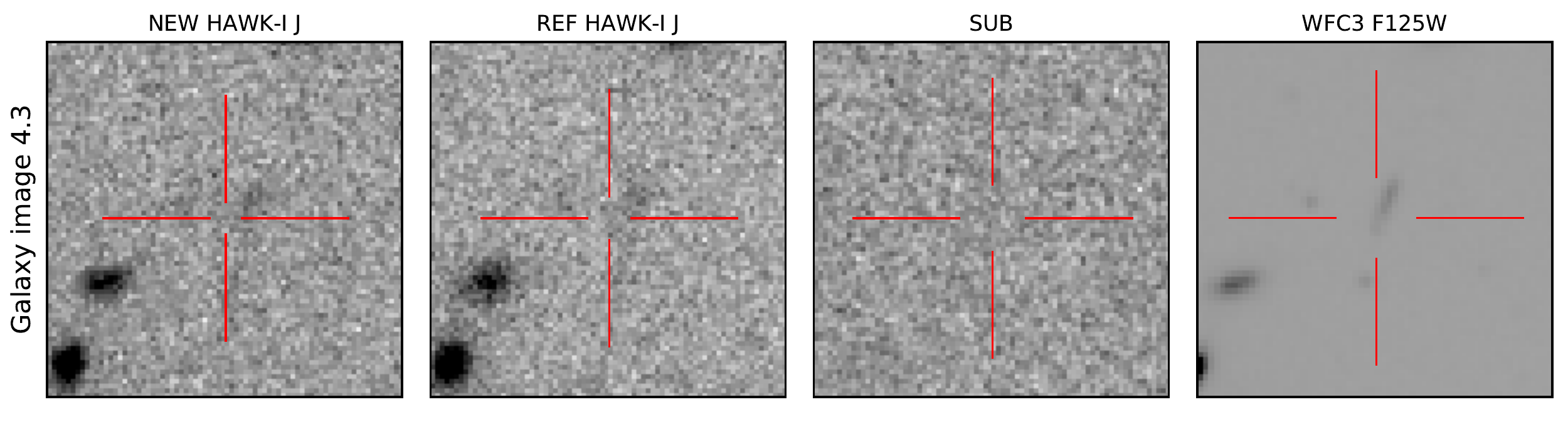}
		\label{f32.pdf}
	\end{center}
\end{figure*}
\begin{figure*} [htbp]
	\begin{center}
		\includegraphics[width=0.8\textwidth]{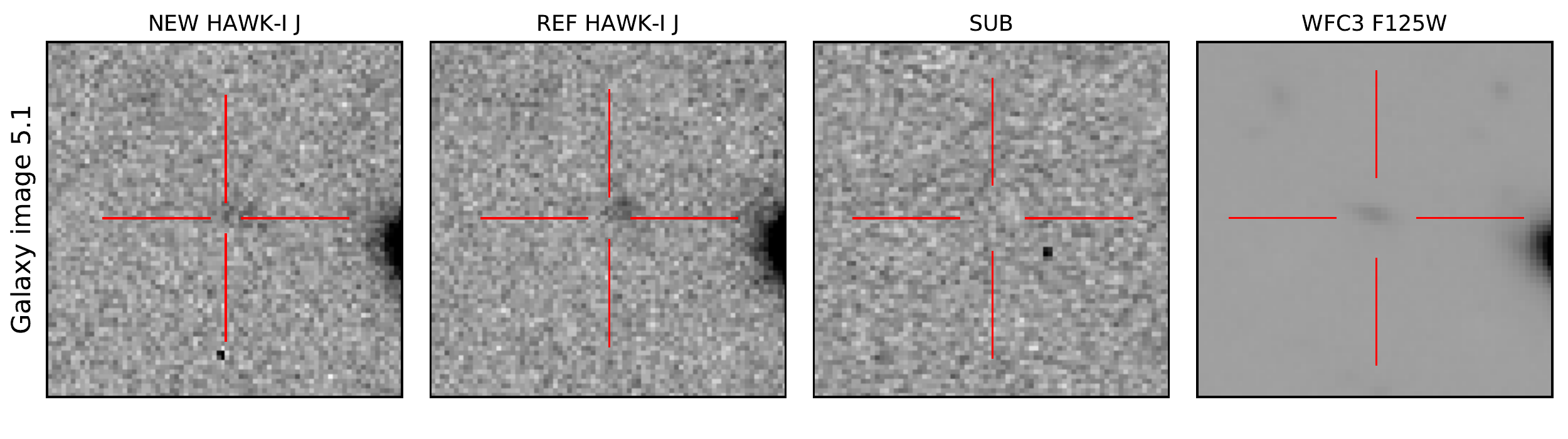}
		\label{f34.pdf}
	\end{center}
\end{figure*}
\begin{figure*} [htbp]
	\begin{center}
		\includegraphics[width=0.8\textwidth]{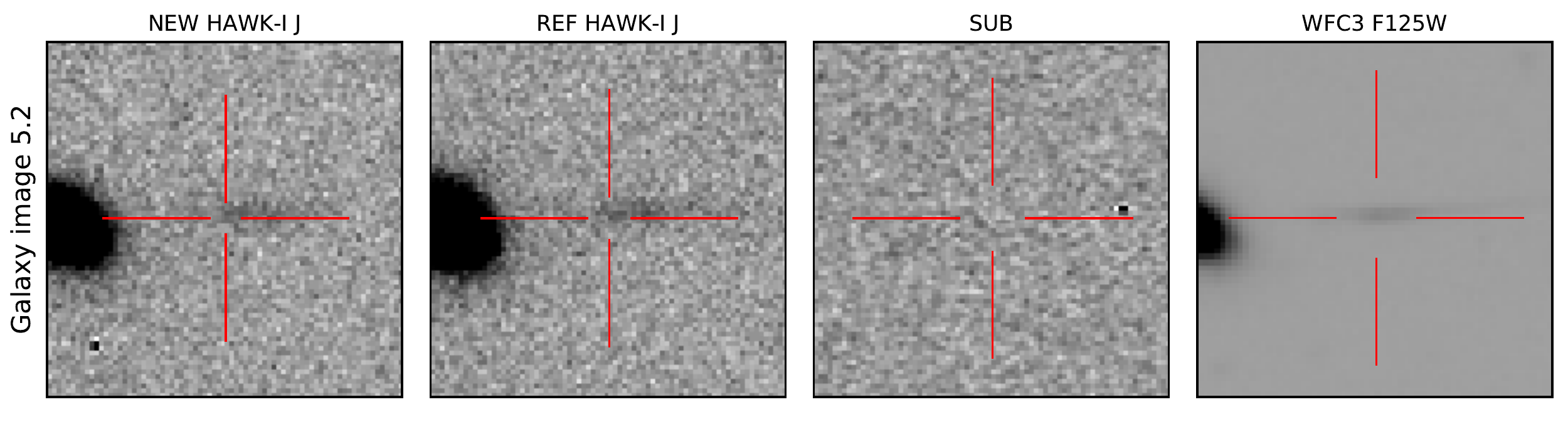}
		\label{f36.pdf}
	\end{center}
\end{figure*}
\begin{figure*} [htbp]
	\begin{center}
		\includegraphics[width=0.8\textwidth]{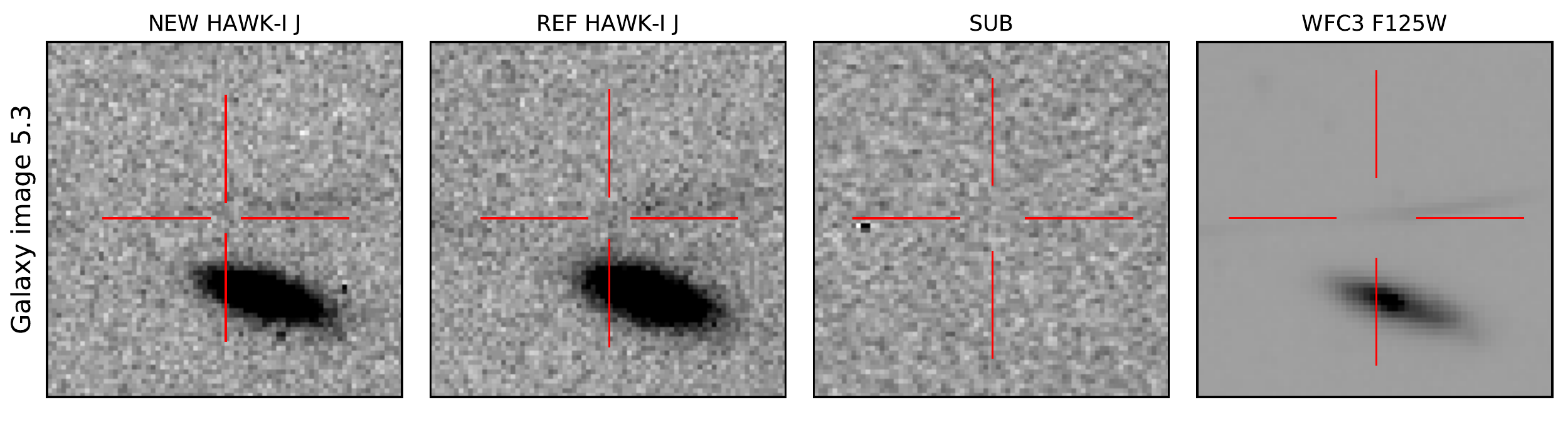}
		\label{f38.pdf}
	\end{center}
\end{figure*}
\begin{figure*} [htbp]
	\begin{center}
		\includegraphics[width=0.8\textwidth]{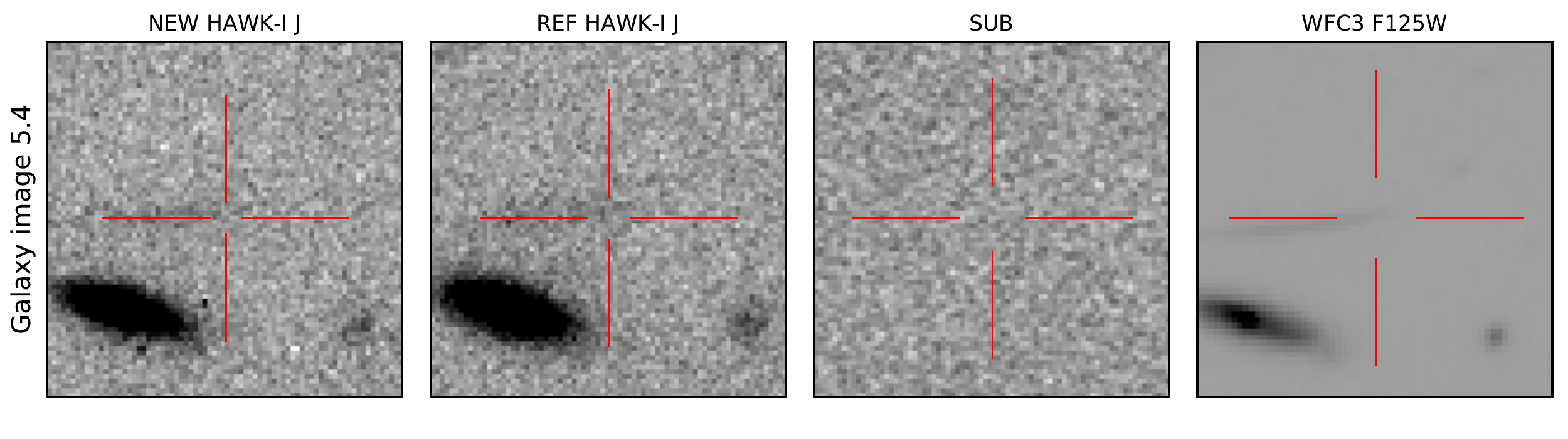}
		\label{f40.pdf}
	\end{center}
\end{figure*}
\begin{figure*} [htbp]
	\begin{center}
		\includegraphics[width=0.8\textwidth]{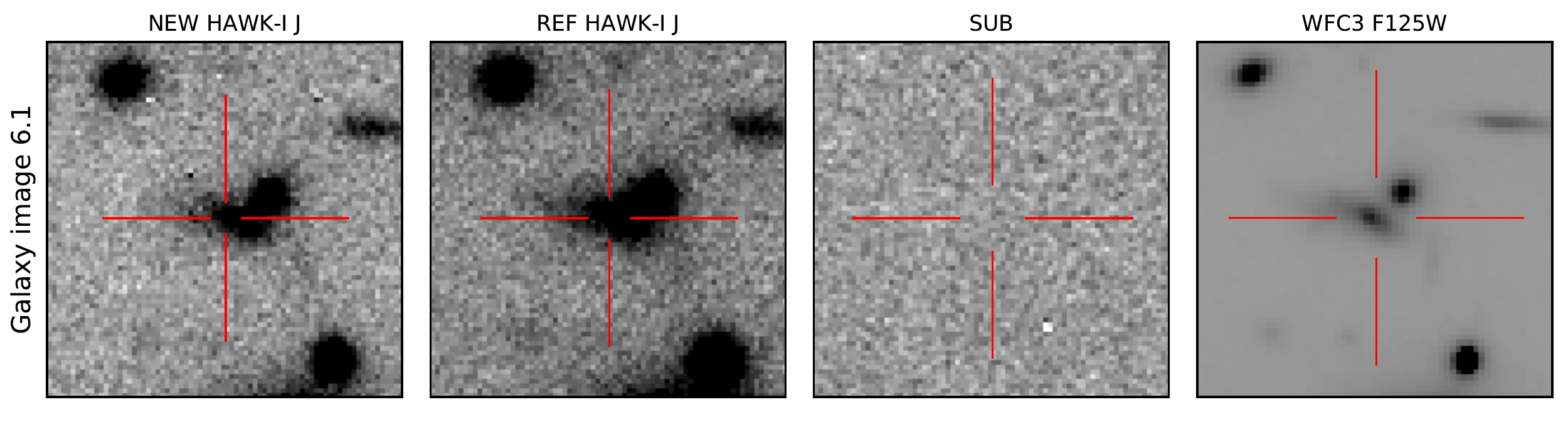}
		\label{f42.pdf}
	\end{center}
\end{figure*}
\begin{figure*} [htbp]
	\begin{center}
		\includegraphics[width=0.8\textwidth]{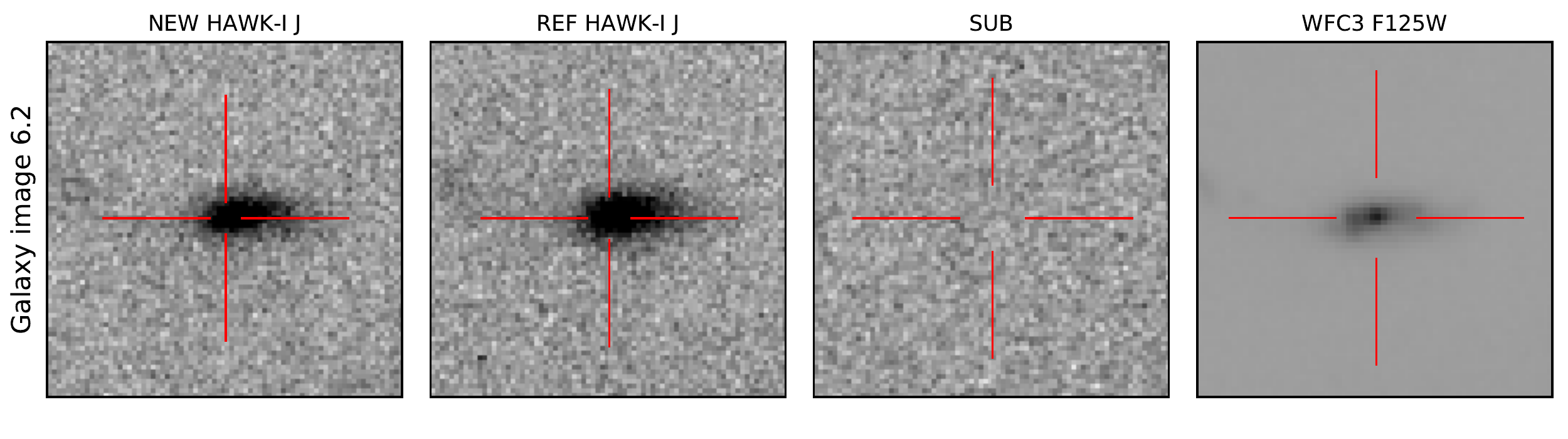}
		\label{f44.pdf}
	\end{center}
\end{figure*}
\begin{figure*} [htbp]
	\begin{center}
		\includegraphics[width=0.8\textwidth]{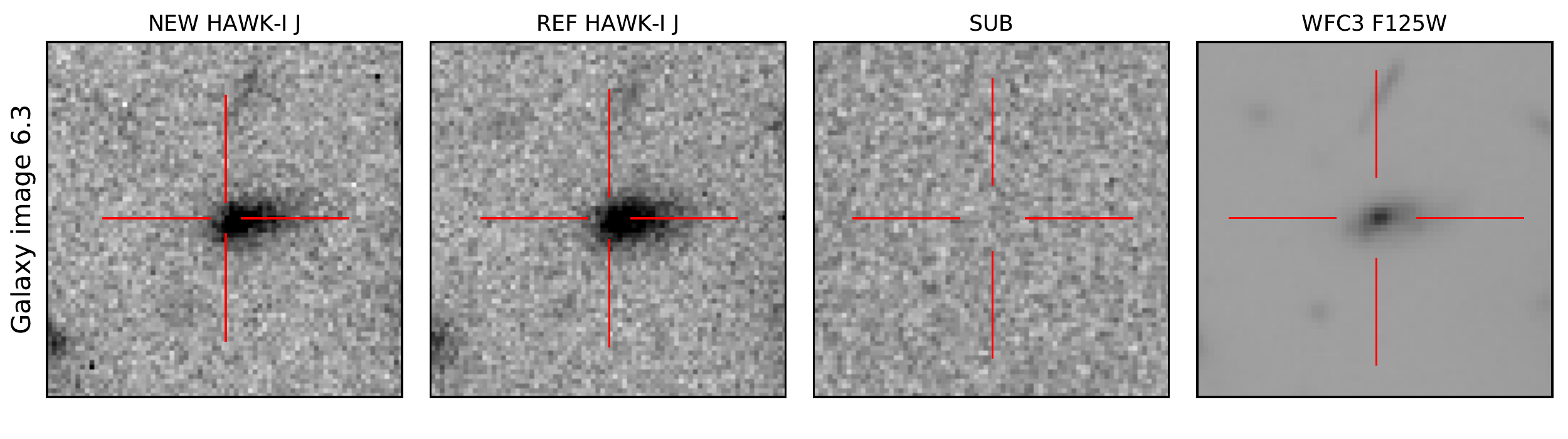}
		\label{f46.pdf}
	\end{center}
\end{figure*}
\begin{figure*} [htbp]
	\begin{center}
		\includegraphics[width=0.8\textwidth]{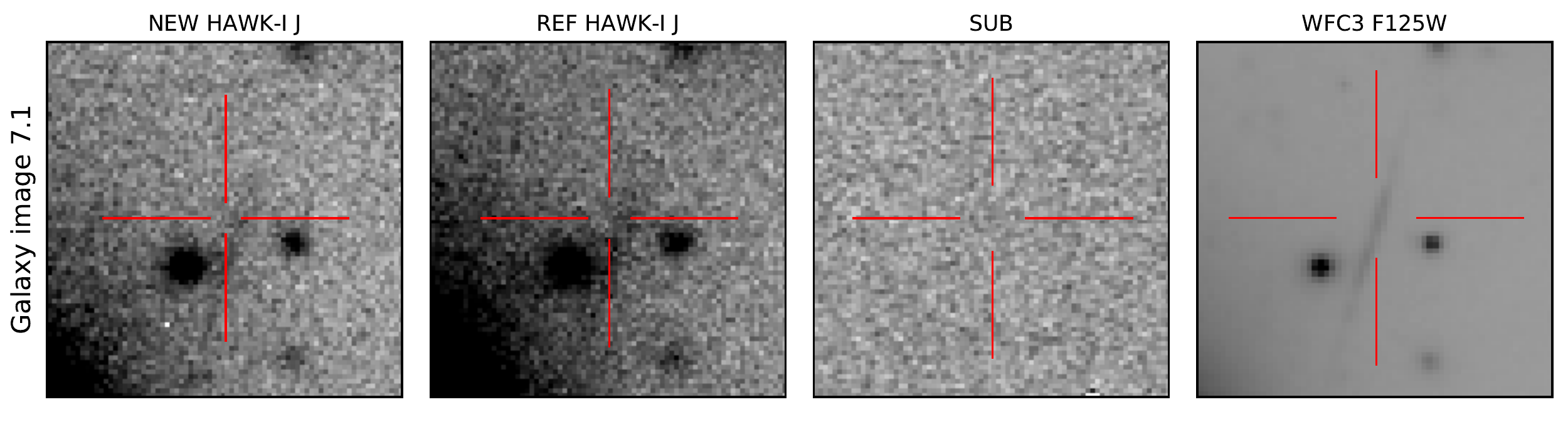}
		\label{f48.pdf}
	\end{center}
\end{figure*}
\begin{figure*} [htbp]
	\begin{center}
		\includegraphics[width=0.8\textwidth]{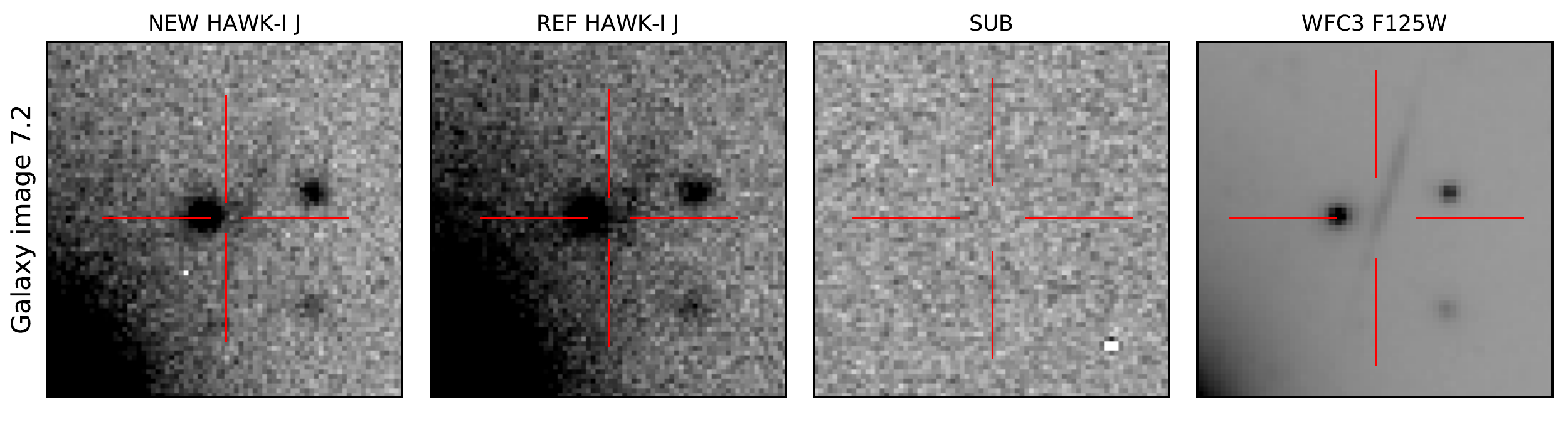}
		\label{f50.pdf}
	\end{center}
\end{figure*}
\begin{figure*} [htbp]
	\begin{center}
		\includegraphics[width=0.8\textwidth]{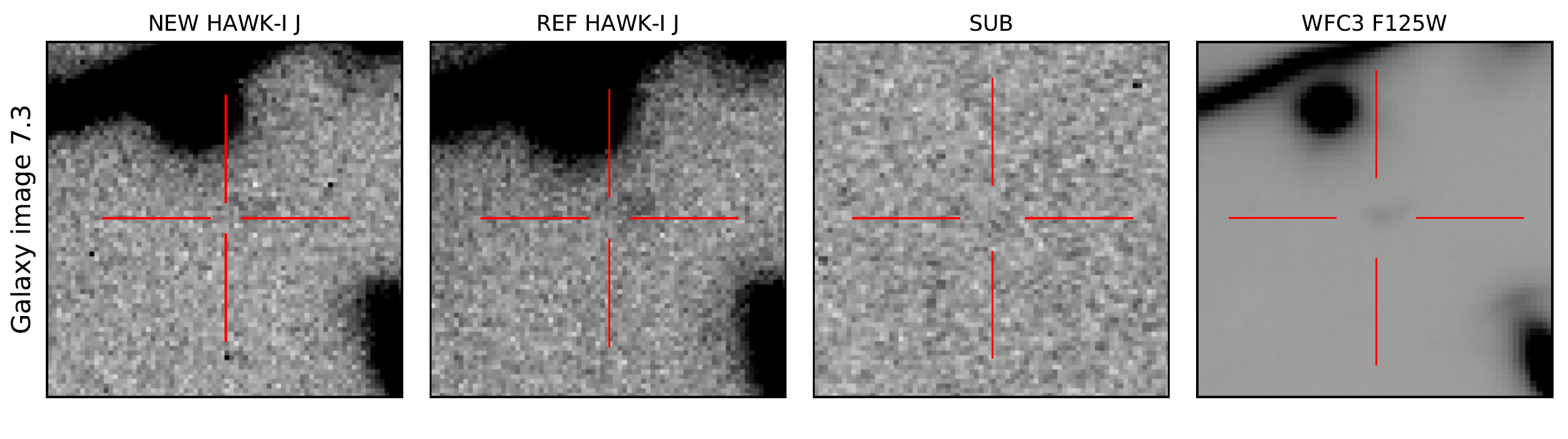}
		\label{f52.pdf}
	\end{center}
\end{figure*}
\begin{figure*} [htbp]
	\begin{center}
		\includegraphics[width=0.8\textwidth]{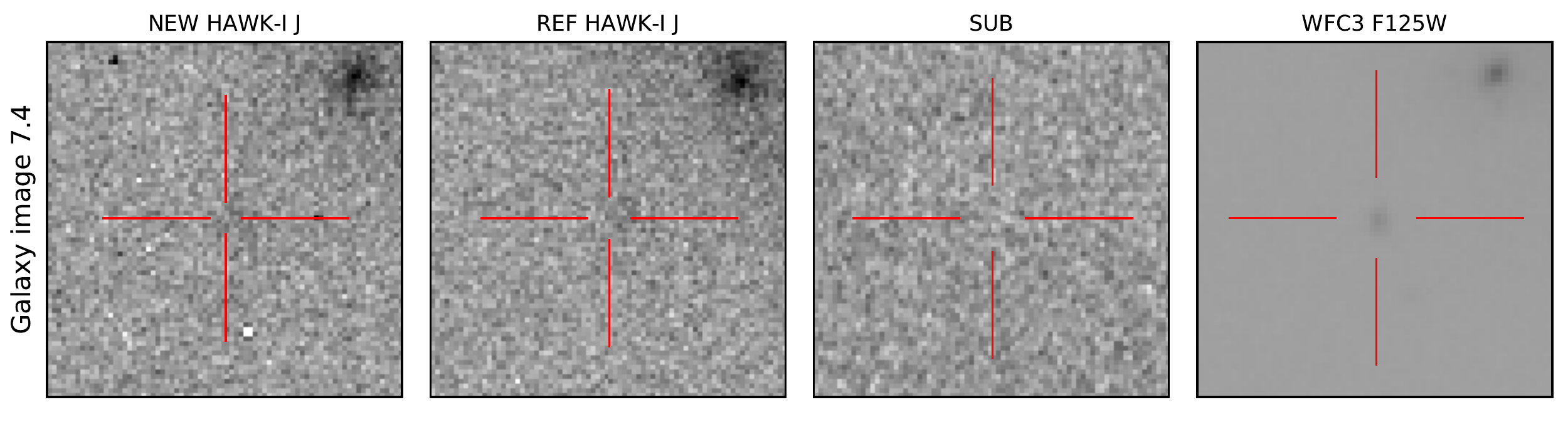}
		\label{f54.pdf}
	\end{center}
\end{figure*}
\begin{figure*} [htbp]
	\begin{center}
		\includegraphics[width=0.8\textwidth]{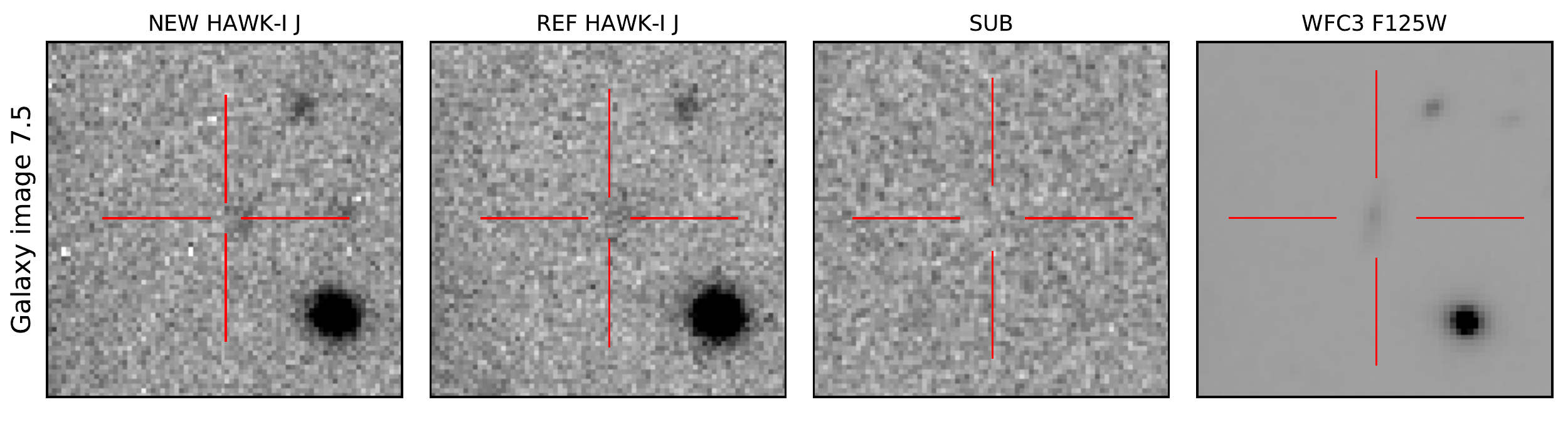}
		\label{f56.pdf}
	\end{center}
\end{figure*}
\begin{figure*} [htbp]
	\begin{center}
		\includegraphics[width=0.8\textwidth]{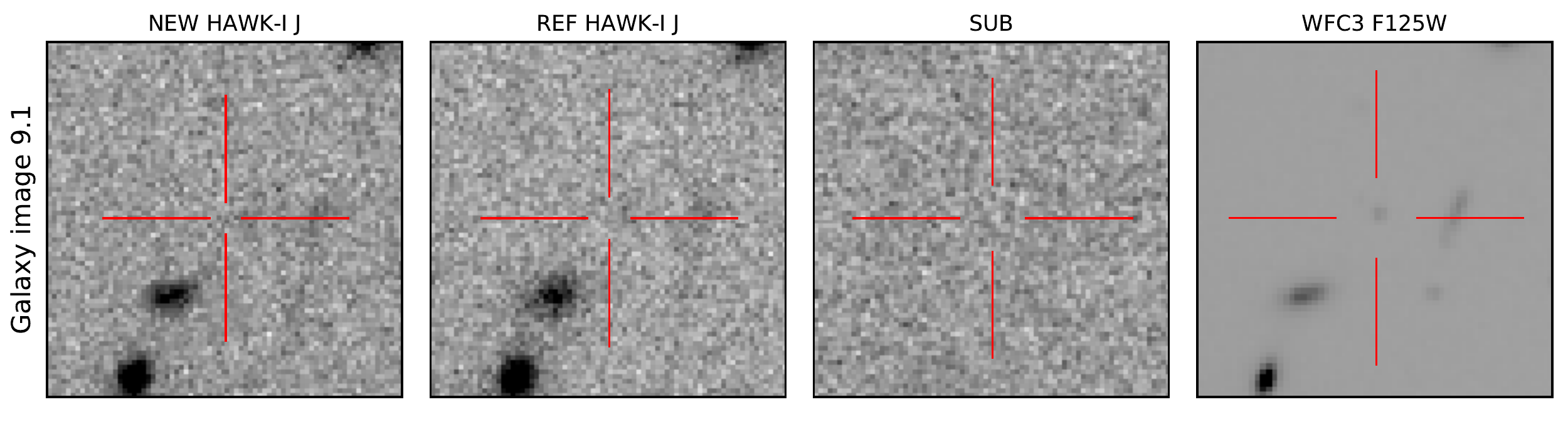}
		\label{f58.pdf}
	\end{center}
\end{figure*}
\begin{figure*} [htbp]
	\begin{center}
		\includegraphics[width=0.8\textwidth]{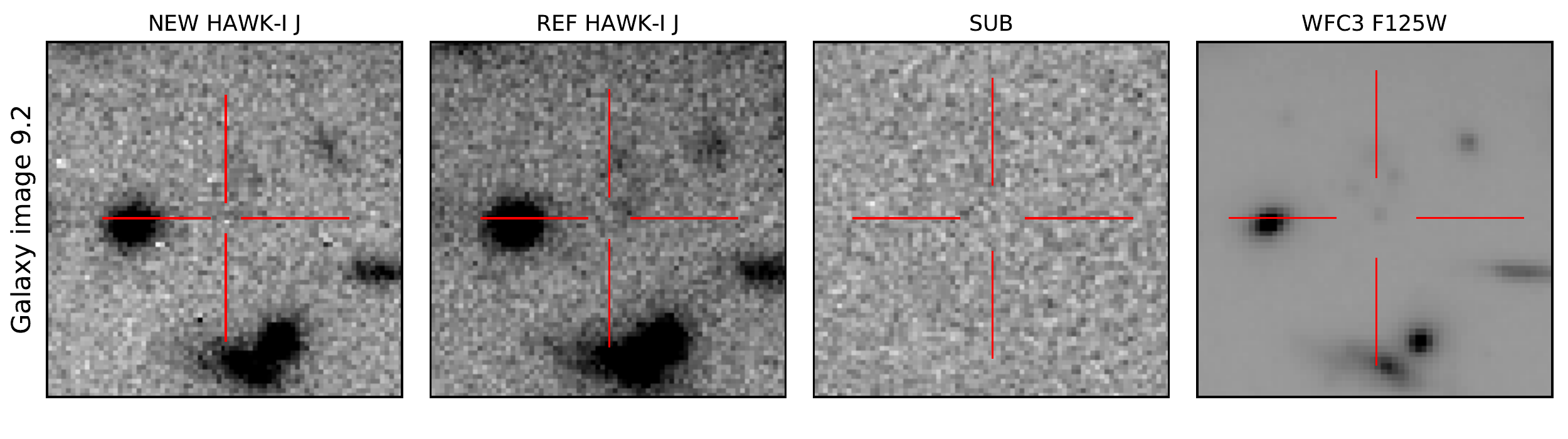}
		\label{f60.pdf}
	\end{center}
\end{figure*}
\begin{figure*} [htbp]
	\begin{center}
		\includegraphics[width=0.8\textwidth]{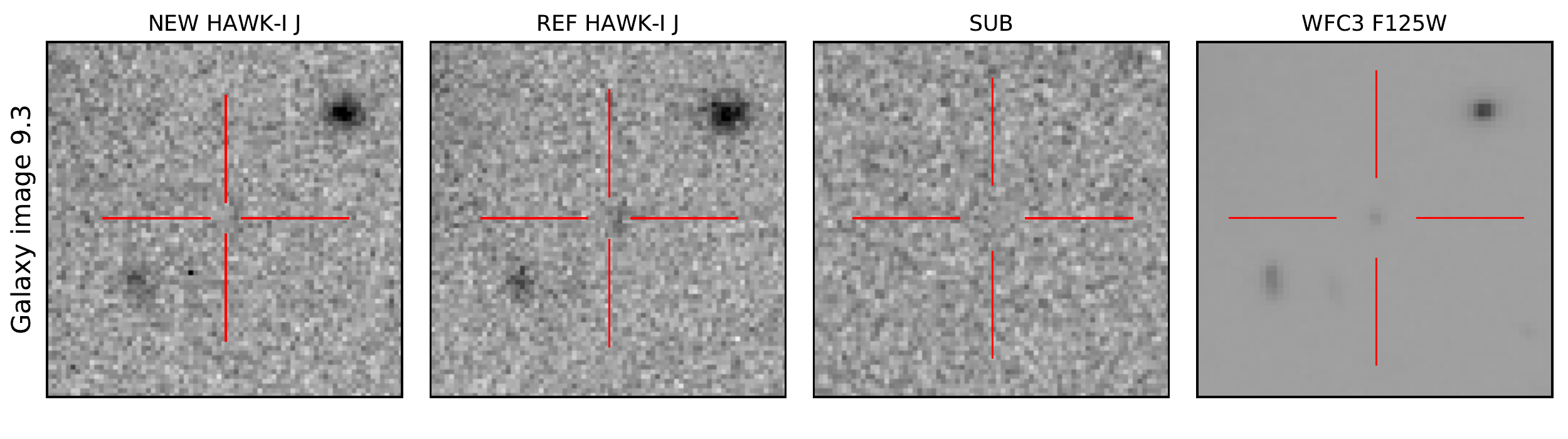}
		\label{f62.pdf}
	\end{center}
\end{figure*}
\begin{figure*} [htbp]
	\begin{center}
		\includegraphics[width=0.8\textwidth]{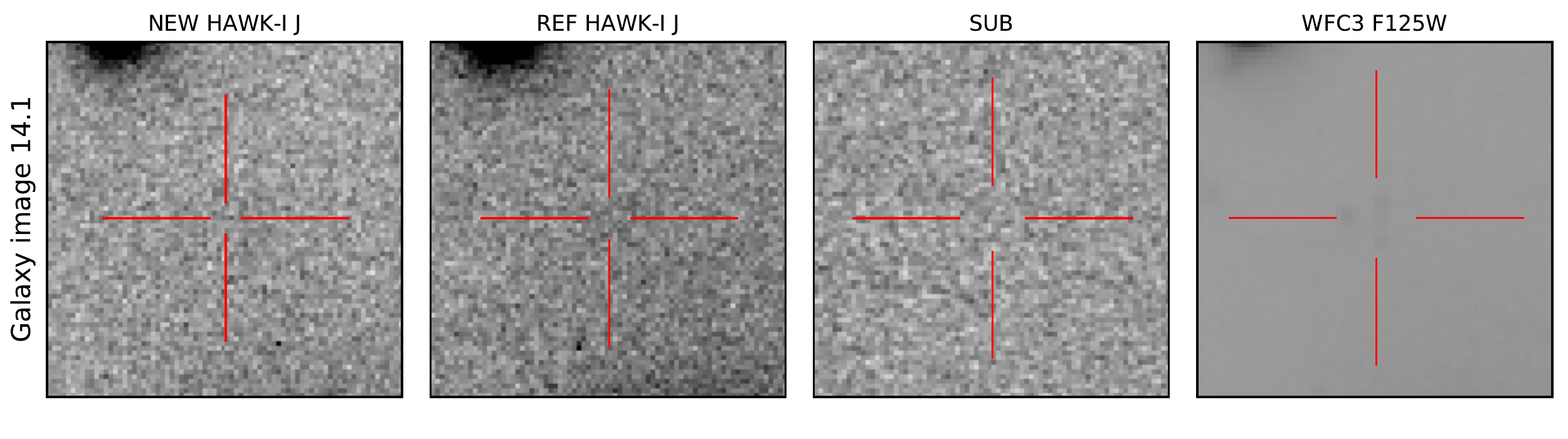}
		\label{f64.pdf}
	\end{center}
\end{figure*}
\begin{figure*} [htbp]
	\begin{center}
		\includegraphics[width=0.8\textwidth]{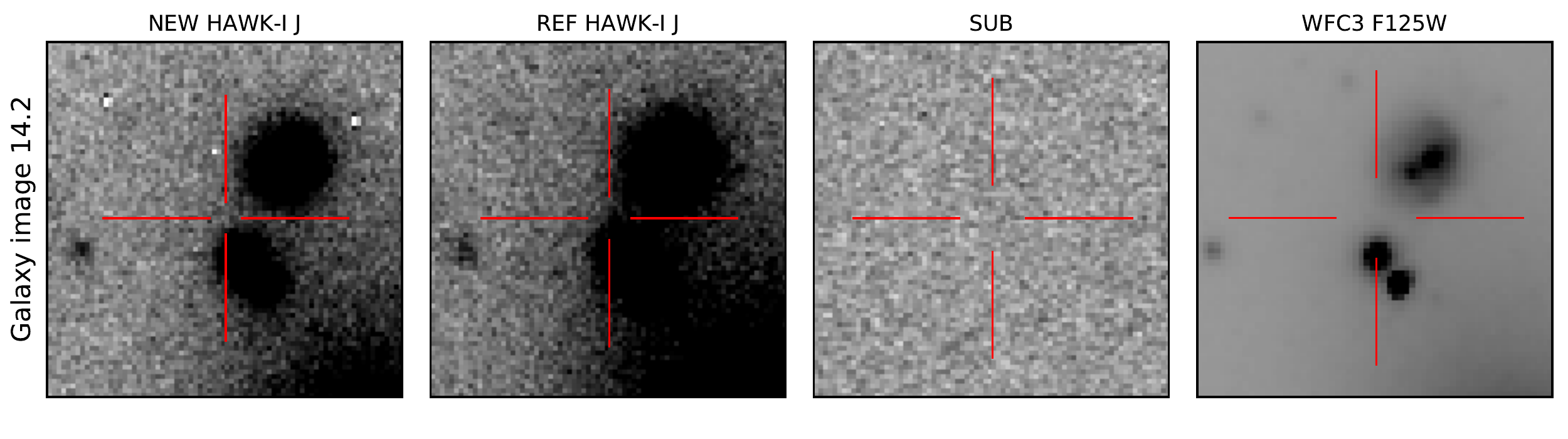}
		\label{f66.pdf}
	\end{center}
\end{figure*}
\begin{figure*} [htbp]
	\begin{center}
		\includegraphics[width=0.8\textwidth]{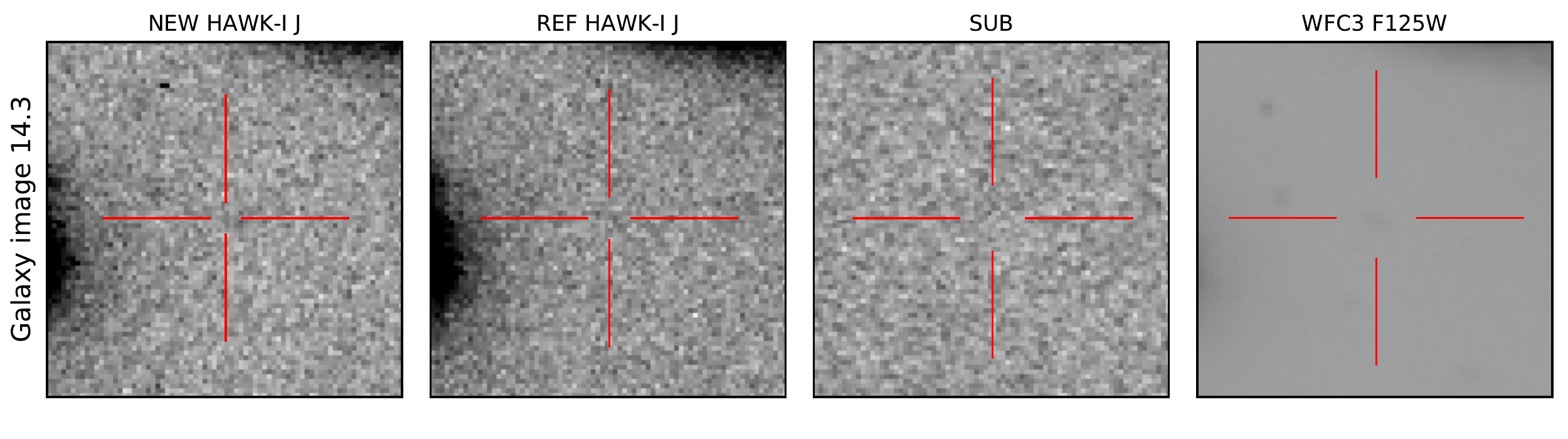}
		\label{f68.pdf}
	\end{center}
\end{figure*}
\begin{figure*} [htbp]
	\begin{center}
		\includegraphics[width=0.8\textwidth]{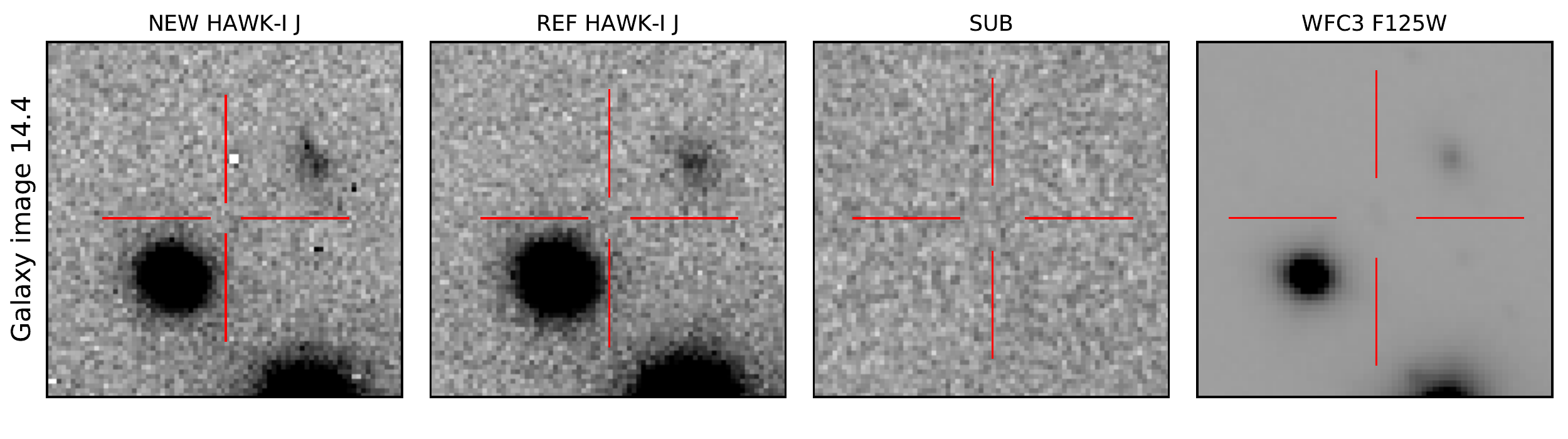}
		\label{f70.pdf}
	\end{center}
\end{figure*}
\begin{figure*} [htbp]
	\begin{center}
		\includegraphics[width=0.8\textwidth]{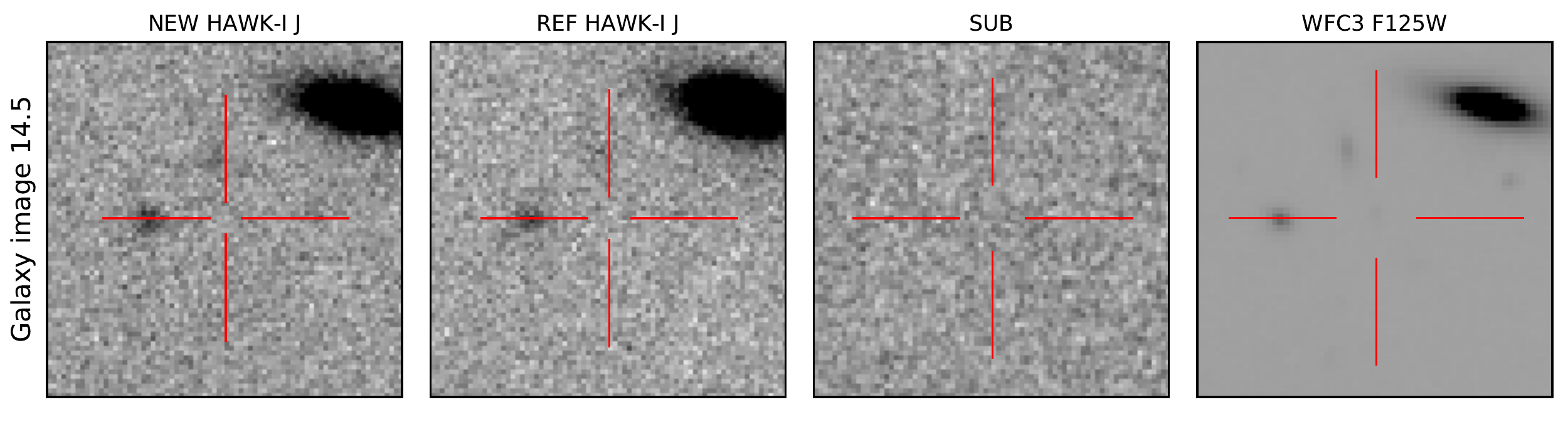}
		\label{f72.pdf}
	\end{center}
\end{figure*}
\begin{figure*} [htbp]
	\begin{center}
		\includegraphics[width=0.8\textwidth]{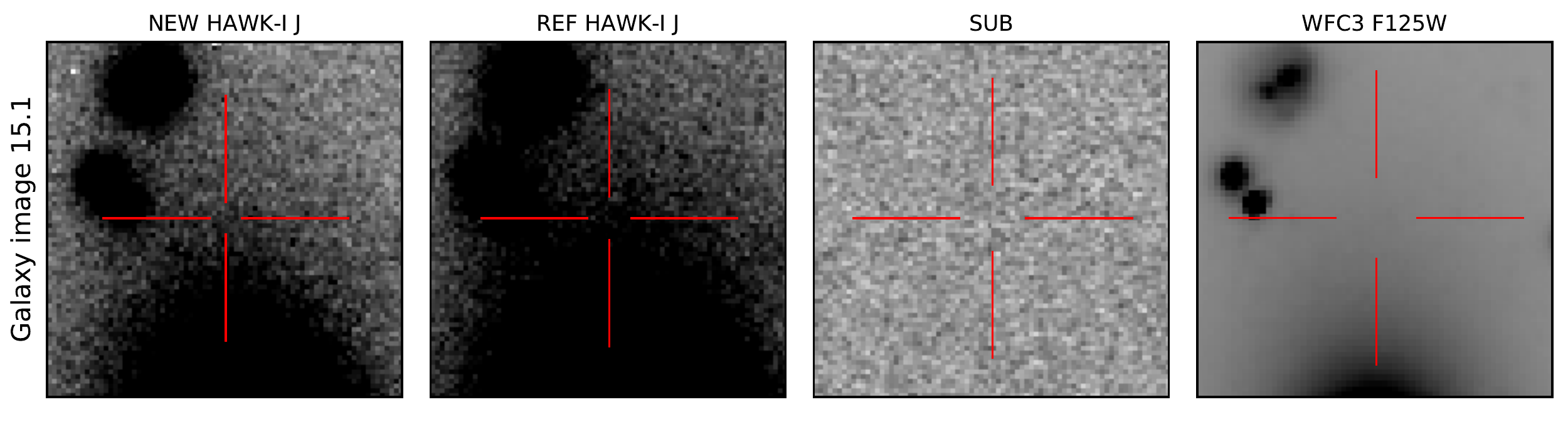}
		\label{f74.pdf}
	\end{center}
\end{figure*}
\begin{figure*} [htbp]
	\begin{center}
		\includegraphics[width=0.8\textwidth]{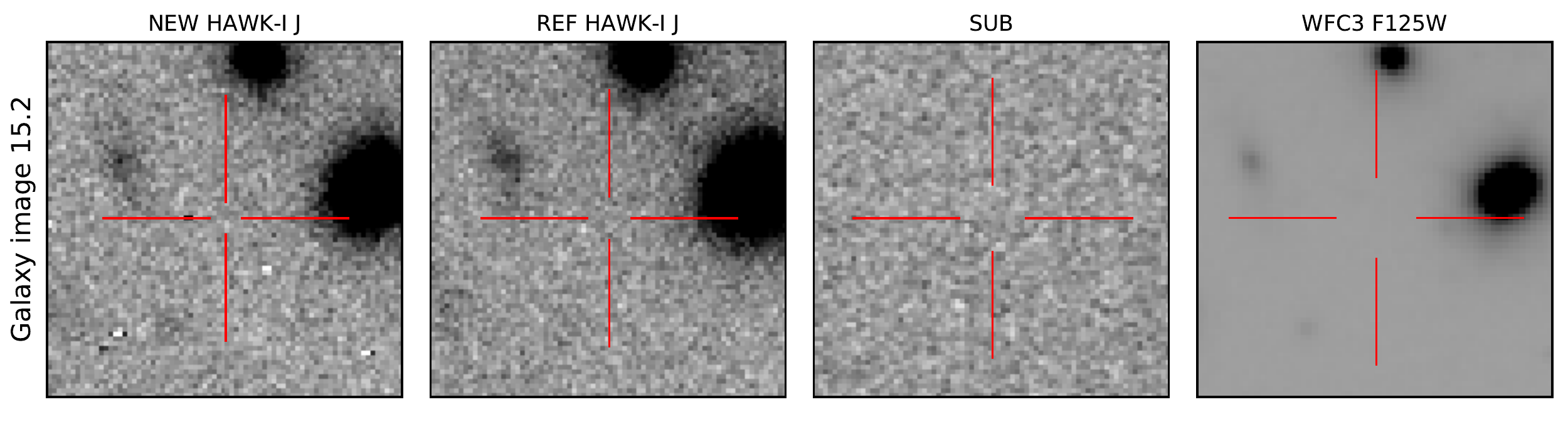}
		\label{f76.pdf}
	\end{center}
\end{figure*}
\begin{figure*} [htbp]
	\begin{center}
		\includegraphics[width=0.8\textwidth]{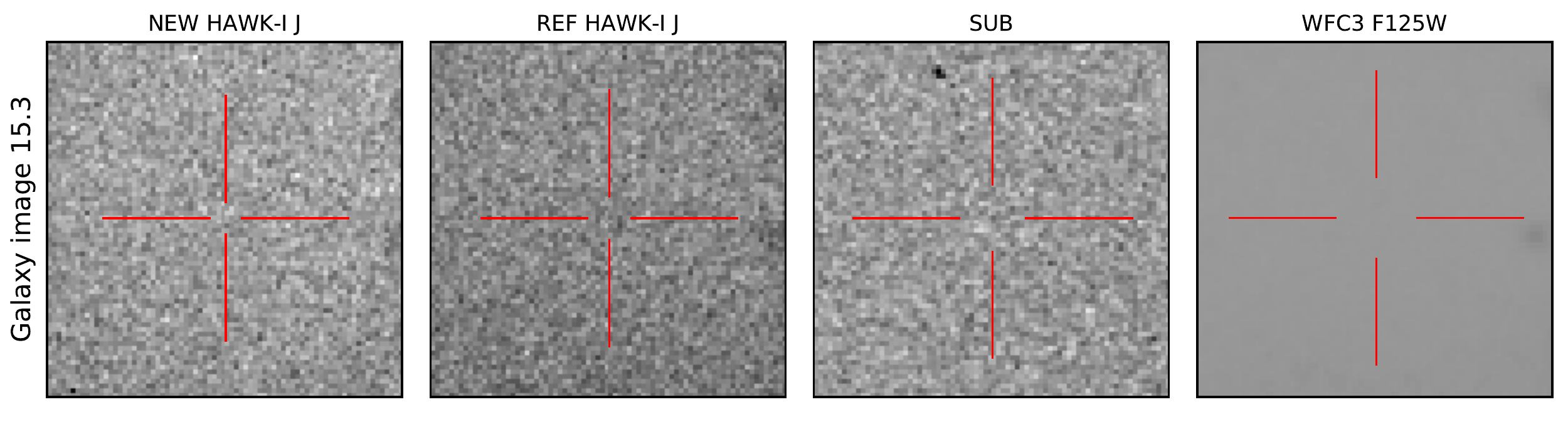}
		\label{f78.pdf}
	\end{center}
\end{figure*}
\begin{figure*} [htbp]
	\begin{center}
		\includegraphics[width=0.8\textwidth]{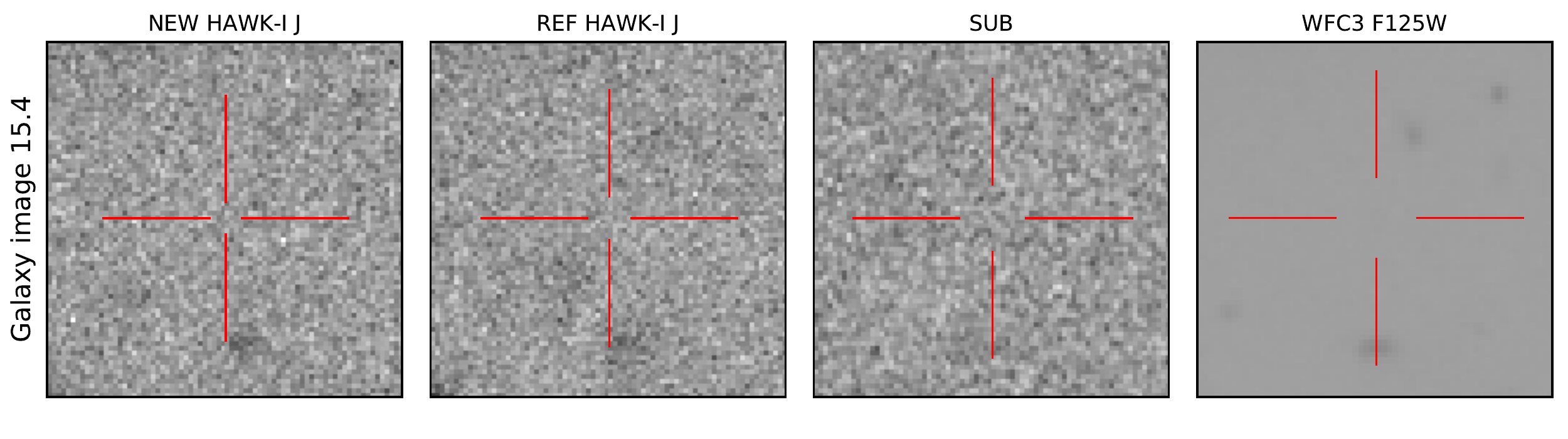}
		\label{f80.pdf}
	\end{center}
\end{figure*}
\begin{figure*} [htbp]
	\begin{center}
		\includegraphics[width=0.8\textwidth]{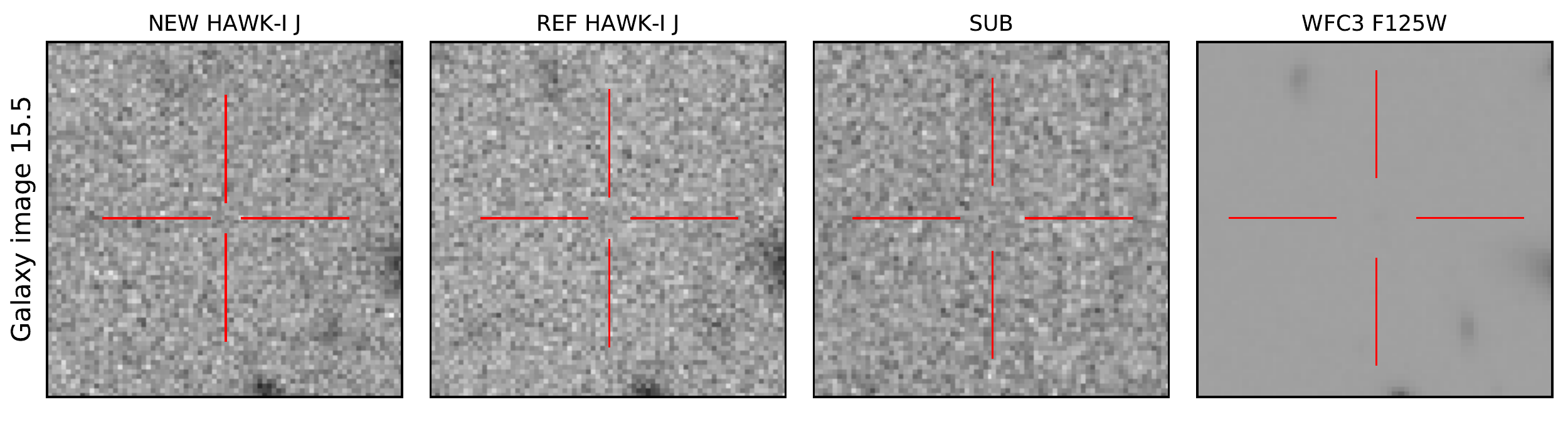}
		\label{f82.pdf}
	\end{center}
\end{figure*}
\begin{figure*} [htbp]
	\begin{center}
		\includegraphics[width=0.8\textwidth]{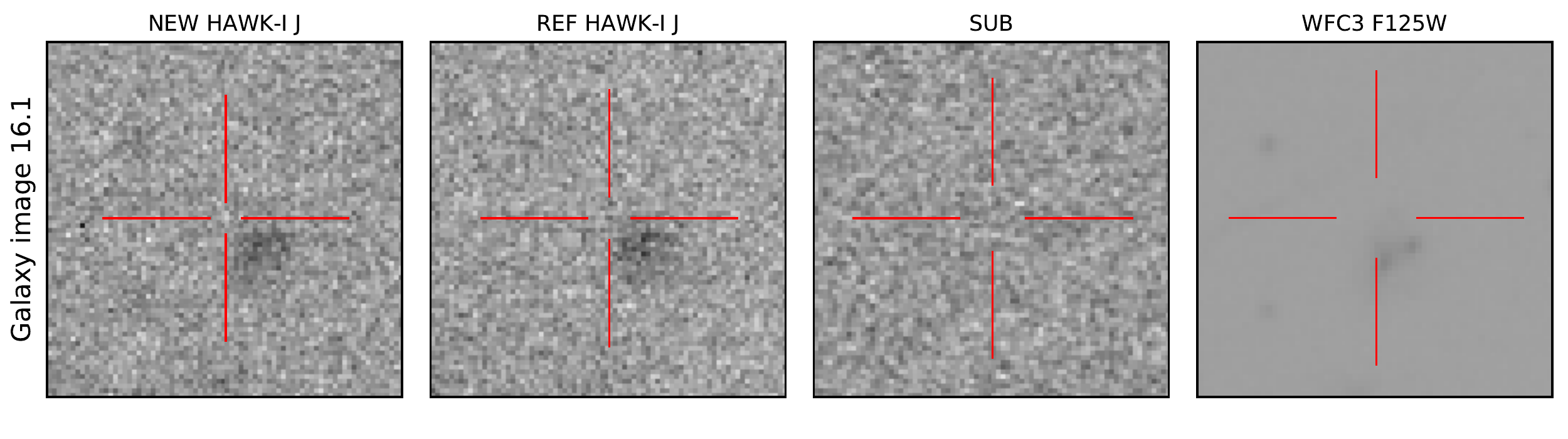}
		\label{f84.pdf}
	\end{center}
\end{figure*}
\begin{figure*} [htbp]
	\begin{center}
		\includegraphics[width=0.8\textwidth]{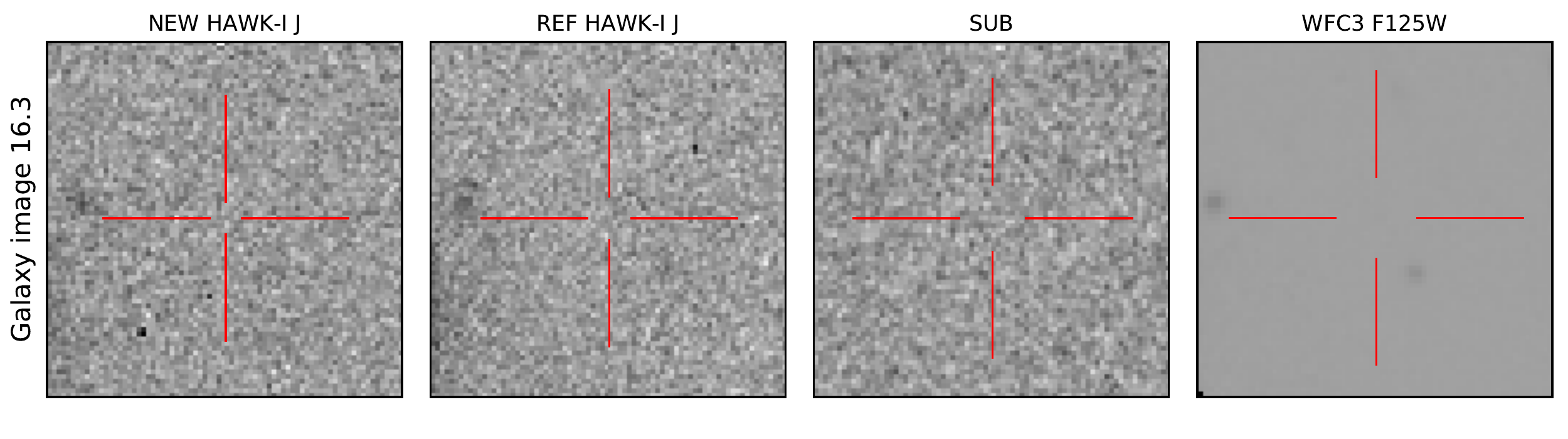}
		\label{f86.pdf}
	\end{center}
\end{figure*}
\begin{figure*} [htbp]
	\begin{center}
		\includegraphics[width=0.8\textwidth]{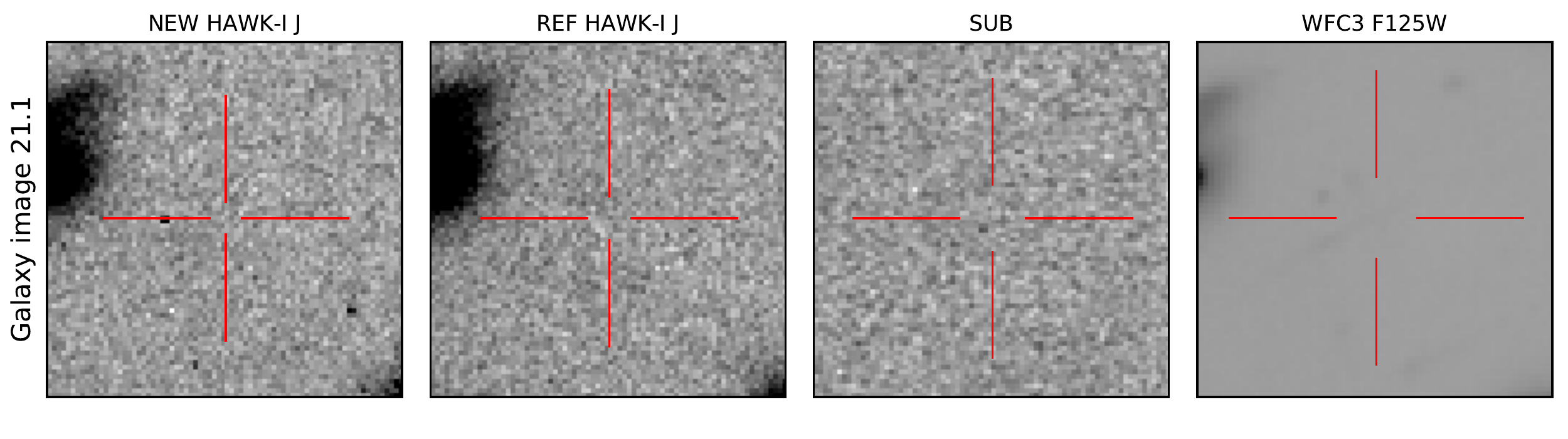}
		\label{f88.pdf}
	\end{center}
\end{figure*}
\begin{figure*} [htbp]
	\begin{center}
		\includegraphics[width=0.8\textwidth]{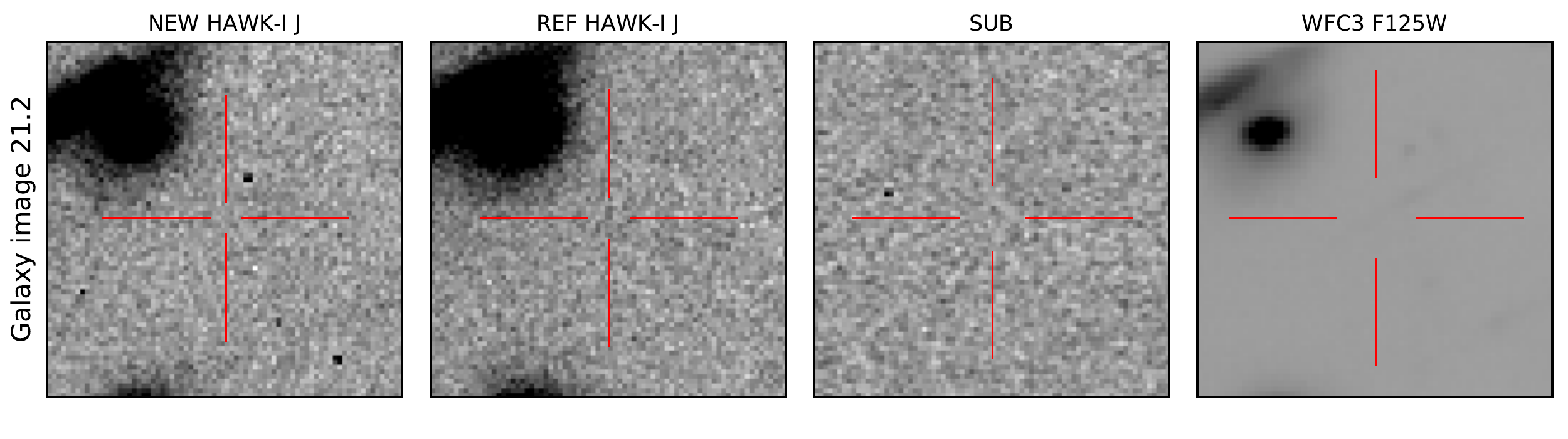}
		\label{f90.pdf}
	\end{center}
\end{figure*}
\begin{figure*} [htbp]
	\begin{center}
		\includegraphics[width=0.8\textwidth]{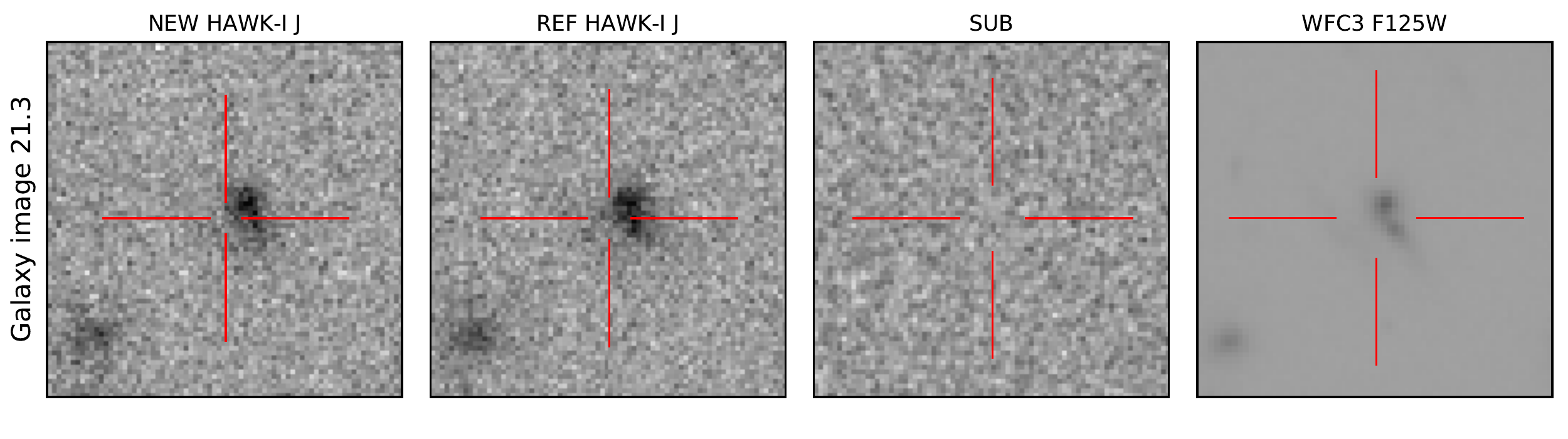}
		\label{f92.pdf}
	\end{center}
\end{figure*}
\begin{figure*} [htbp]
	\begin{center}
		\includegraphics[width=0.8\textwidth]{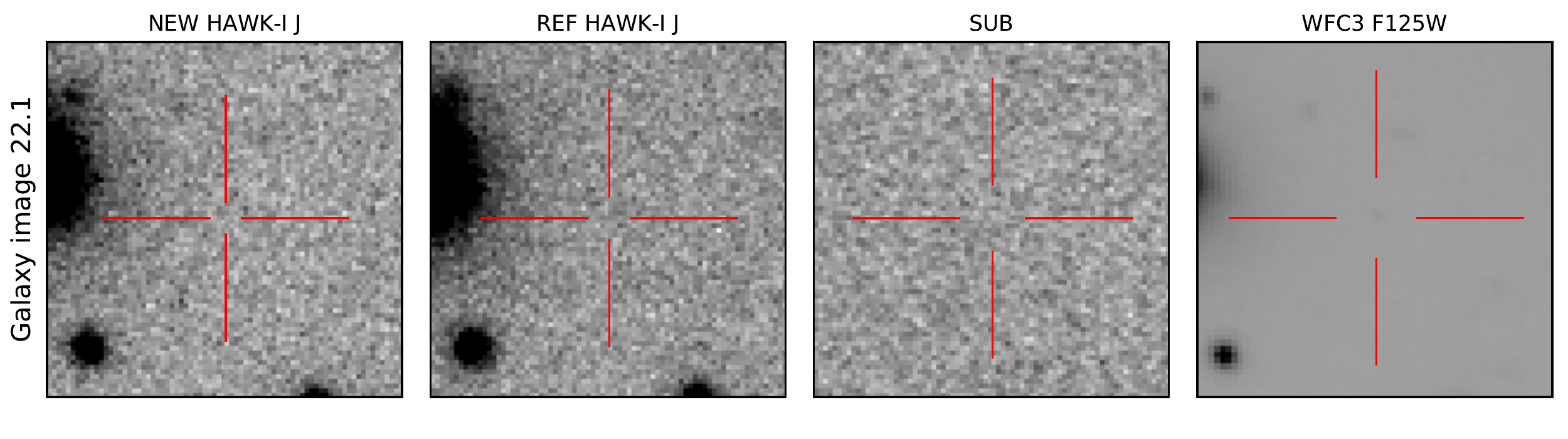}
		\label{f94.pdf}
	\end{center}
\end{figure*}
\begin{figure*} [htbp]
	\begin{center}
		\includegraphics[width=0.8\textwidth]{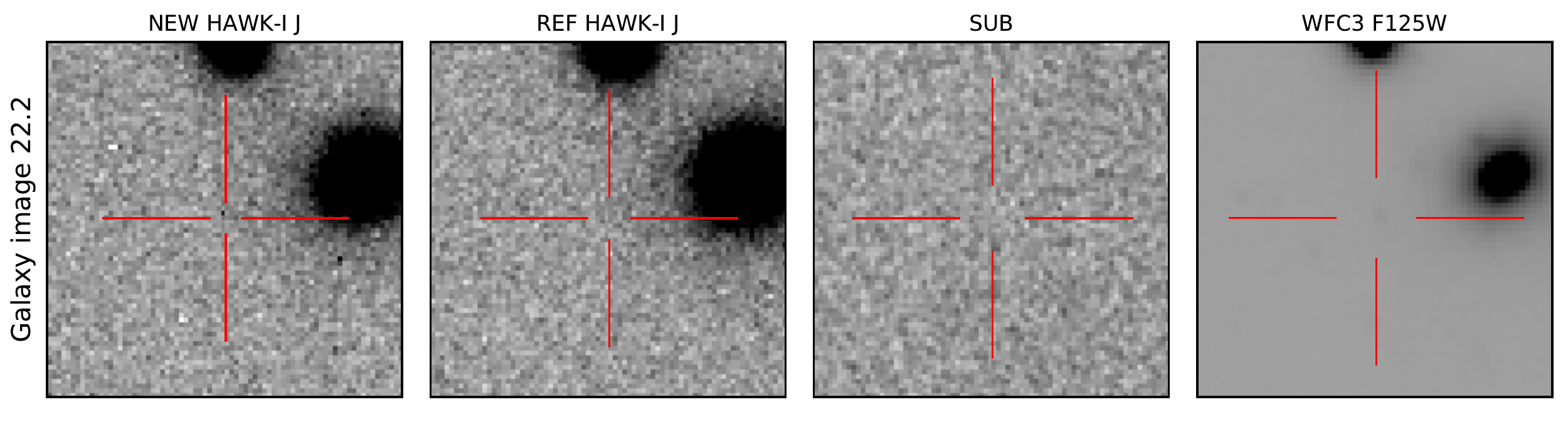}
		\label{f96.pdf}
	\end{center}
\end{figure*}
\begin{figure*} [htbp]
	\begin{center}
		\includegraphics[width=0.8\textwidth]{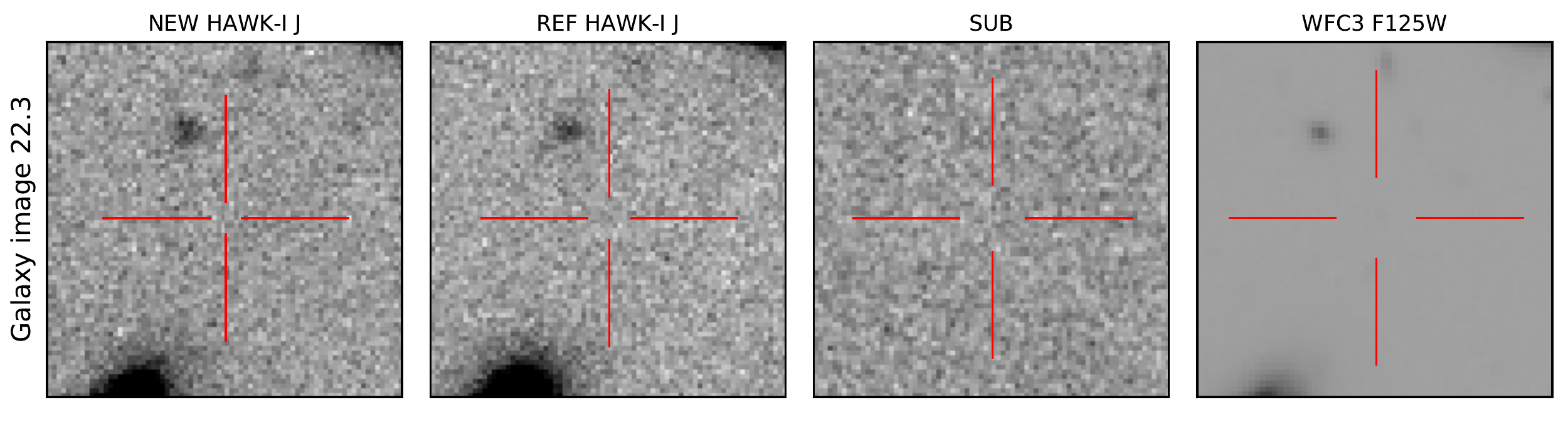}
		\label{f98.pdf}
	\end{center}
\end{figure*}

\end{document}